\documentclass[11pt,pdftex]{article}
\usepackage{times}
\usepackage{amssymb,amsmath}
\usepackage{amsthm}

\usepackage[pdftex]{graphicx}
\usepackage{multirow}
\usepackage{caption}
\usepackage[T1]{fontenc}
\usepackage{calc}
\usepackage{setspace}
\usepackage{color,colortbl}
\usepackage{url}

\usepackage{rotfloat}
\usepackage{url}
\usepackage{rotating}
\usepackage{hyperref}
\usepackage{subfig}
\usepackage{longtable}
\usepackage{lscape}
\usepackage{multirow}
\usepackage[toc,page]{appendix}
\usepackage[english]{babel}
\addto{\captionsenglish}{\renewcommand{\appendixname}{Supplementary Information}}
\renewcommand{\appendixname}{Supplementary Information}

\newcommand{\newsubref}{\protect\subref}
\catcode`\_=11                %%% this small piece of codes %%%%
\edef\ImageUnderscore{_}      %%% is to recoganize the      %%%%
\catcode`\_=8                 %%% the underscores in image  %%%%
\def\Image[#1]#2{{%           %%% files, so that one does   %%%%
\let\_=\ImageUnderscore       %%% not need to add special   %%%%
\includegraphicsII[#1]{#2}}}  %%% characters. @author QZ    %%%%
\begin{document}

%
%\bibliographystyle{nature}
%

%\draft

\title{Characterizing scientific production and consumption in Physics}

\author{ Qian Zhang$^{1}$, Nicola Perra$^{1}$, Bruno Gon\c calves$^{2}$, Fabio Ciulla$^{1}$, Alessandro Vespignani$^{1,3,4}$\footnote{To whom correspondence
should be addressed; email: a.vespignani@neu.edu}}
\date{ }
\maketitle

\begin{center}
\small{
${}^{1}$Laboratory for the Modeling of Biological and Socio-technical Systems, Northeastern University, Boston MA 02115 USA\\
    ${}^{2}$Aix Marseille Universit\'e, CNRS, CPT, UMR 7332, 13288 Marseille, France\\
    ${}^{3}$Institute for Scientific Interchange Foundation, Turin 10133, Italy\\
    ${}^{4}$Institute for Quantitative Social Sciences at Harvard University, Cambridge, MA,02138}
\end{center}

\begin{abstract}
\bf
We analyze the entire publication database of the American Physical Society generating longitudinal (50 years) citation networks geolocalized at the level of single urban areas. We define the knowledge diffusion proxy, and scientific production ranking algorithms to capture the spatio-temporal dynamics of Physics knowledge worldwide. By using the knowledge diffusion proxy we identify the key cities in the production and consumption of knowledge in Physics as a function of time. The results from the scientific production ranking algorithm allow us to characterize the top cities for scholarly research in Physics. Although we focus on a single dataset concerning a specific field, the methodology presented here opens the path to comparative studies of the dynamics of knowledge across disciplines and research areas.
\end{abstract}

\maketitle
Over the last decade, the digitalization of publication datasets has propelled bibliographic studies allowing for the first time access to the geospatial distribution of millions of publications, and citations at different granularities~\cite{Narin1975,Frame1977,May1997,Batty03,Leydesdorff2005,Horta2007,Fortunato2012,Boner2013} (see~\cite{Frenken2009} for a review). More precisely, authors' name, affiliations, addresses, and references can be aggregated at different scales, and used to characterize publications and citations patterns of single papers~\cite{Redner1998,Chen2007}, journals~\cite{Garfield1972,Bergstrom2007}, authors~\cite{Hirsch2005,Egghe2006,Hirsch2007}, institutions~\cite{Borner2006}, cities~\cite{Bornmann2011}, or countries~\cite{King2004}.
The sheer size of the datasets allows also system level analysis on research production and consumption~\cite{Adams2012}, migration of authors~\cite{Laudel2003,VanNoorden2012}, and change in production in several regions of the world as a function of time~\cite{Leydesdorff2005,Horta2007}, just to name a few examples. At the same time those analyses have spurred an intense research activity aimed at defining metrics able to capture the importance/ranking of  authors, institutions,  or even entire countries~\cite{Garfield1979,Egghe1990,Hirsch2005,Egghe2006,Borner2006,Kleinberg1999,CiteRank,Castillo2007,Sidiropoulos2007,SARA}. Whereas such large datasets are extremely useful in understanding scholarly networks and in charting the creation of knowledge, they are also pointing out the limits of our conceptual and modeling frameworks \cite{scharnhorst2012} and call for a deeper understanding of the dynamics ruling the diffusion and fruition of knowledge across the the social and geographical space. \\

In this paper we study citation patterns of articles published in the American Physical Society (APS) journals in a fifty-year time interval ($1960$-$2009$)~\cite{aps10-1}. Although in the early years of this period the dataset was obviously biased toward the scholarly activity within the USA, in the last twenty years only about 35\% of the papers are produced in the USA. The same amount of production has been observed in databases that include multiple journals, and disciplines\cite{King2004,Fortunato2012}. Indeed  the journals of the APS are considered worldwide as reference publication venues that well represent the international research activity in Physics. Furthermore this dataset does not bundle different disciplines and publication languages, providing a homogeneous dataset concerning Physics scholarly research. For each paper we geolocalize the institutions contained in the authors' affiliations. In this way we are able to associate each paper in the database  with specific urban areas. This defines a time resolved, geolocalized citation network including 2,307 cities around the world engaged in the production of scholarly work in the area of Physics. Following previous works~\cite{Borner2006,Boner2013} we assume that the number of given or received citations is a proxy of knowledge consumption or production, respectively. More precisely, we assume that citations are the currency traded between parties in the knowledge exchange. Nodes that receive citations export their knowledge to others. Nodes that cite other works, import knowledge from others. According to this assumption we classify nodes considering the unbalance in their trade. Knowledge producers are nodes that are cited (export) more than they cite, (import). On the contrary, we label as consumers nodes that cite (import) more than they are cited (export). Using this classification, we define the knowledge diffusion proxy algorithm to explore how scientific knowledge flows from producers to consumers. This tool explicitly assumes a systemic perspective of knowledge diffusion, highlighting the global structure of scientific production and consumption in Physics.\\

The temporal analysis reveals interesting patterns and the progressive delocalization of knowledge producers. In particular, we find that in the last twenty years 
the geographical distribution of knowledge production has drastically changed. A paramount example is the transition in the USA from a knowledge production localized around major urban areas in the east and west coast to a broad geographical distribution where a significant part of the knowledge production is now occurring also in the midwestern and southern states in USA. Analogously, we observe the early 90s dominance of UK and Northern Europe to subside to an increase of production from  France, Italy and several regions of Spain. Interestingly, the last decade shows that several of China's urban areas are emerging as the largest knowledge consumers worldwide. The reasons underlying this phenomenon may be related to the significant growth of the economy and the research/development compartment in China in the early 21${}^{\mathrm{th}}$ century \cite{WorldBankData}. This positive stimulus, pushed up also the scientific consumption with a large number of paper citing work from other world areas. Indeed, the increase of publications is associated to an increase of the citations unbalance, moving China to the top rank as consumers since the recent influx of its new papers has not yet had the time to accumulate citations. \\

Although the knowledge diffusion proxy provides a measure of knowledge production and consumption, it may be inadequate in providing a rank of the most authoritative cities for Physics research. Indeed, a key issue in appropriately ranking the knowledge production, is that not all citations have the same weight. Citations coming from authoritative nodes are \emph{heavier} than others coming from less important nodes, thus defining a recursive diffusion of ranking of nodes in the citation network. In order to include this element in the ranking of cities we propose the scientific production ranking algorithm. This tool, inspired by the PageRank~\cite{brin98-1}, allows us to define the rank of each node, as function of time, going beyond the knowledge diffusion proxy or simple local measures as citation counts or h-index~\cite{Hirsch2005}. In this algorithm the importance of each node diffuses through the citation links. The rank of a node is determined by the rank of the nodes that cite it, recursively, thus implicitly weighting differently citations from highly (lowly) ranked nodes. Also in this case we observe noticeable changes in the ranking of cities along the years. For instance the presence of both European and Asian cities in the top $100$ list increases by $50\%$ in the last $20$ years. This findings suggest that the Internet, digitalization and accessibility of publications are creating a more levelled playing field where the dominance of specific area of the world is being progressively eroded to the advantage of a more widespread and complex knowledge production and consumption dynamic.\\
  
\section*{Results} 
We focus our analysis on the APS dataset~\cite{aps10-1}. It contains all the papers published by the APS from $1893$ to $2009$. We consider only the last $50$ years due to the incomplete geolocalization information available for the early years. During this period, the large majority of indexed papers, $97.47\%$, contain complete information such as authors name, journal of publication, day of publication, list of affiliations and list of citations to other articles published in APS journals.  We geolocalized $96.97\%$ of papers at urban area level with an accuracy of $98.5\%$. We refer the reader to the Methods section and to the Supplementary Information (SI) for the detailed description of the dataset and the techniques developed to geolocalize the affiliations.\\

In total, only $43\%$ of papers has been produced inside the USA. Interestingly, over time this fraction has decreased. For example, in the $60$'s it was $85.59\%$, while in the last $10$ years decreased to just $36.67\%$. While one might assume that the APS dataset is biased toward the USA scientific community, the percentage of publications contributed by the USA in APS journals after $1990$ is almost the same as in other publication datasets \cite{King2004,Fortunato2012}. These alternative datasets contain journals published all over the world and mix different scientific disciplines. This supports the idea that the APS journals are now attracting the worldwide physics scientific community independently of nationality, and fairly represent the world production and consumption of Physics. It is not possible to provide quantitative analysis of possible nationality bias and disentangle it by an actual change of the dynamic of knowledge production. For this reason, and in order to minimize any bias in the analysis we focus our analysis in the last $20$ years of data.\\

\begin{figure}[ht]\centering
\includegraphics[width=0.54\columnwidth,angle=0]{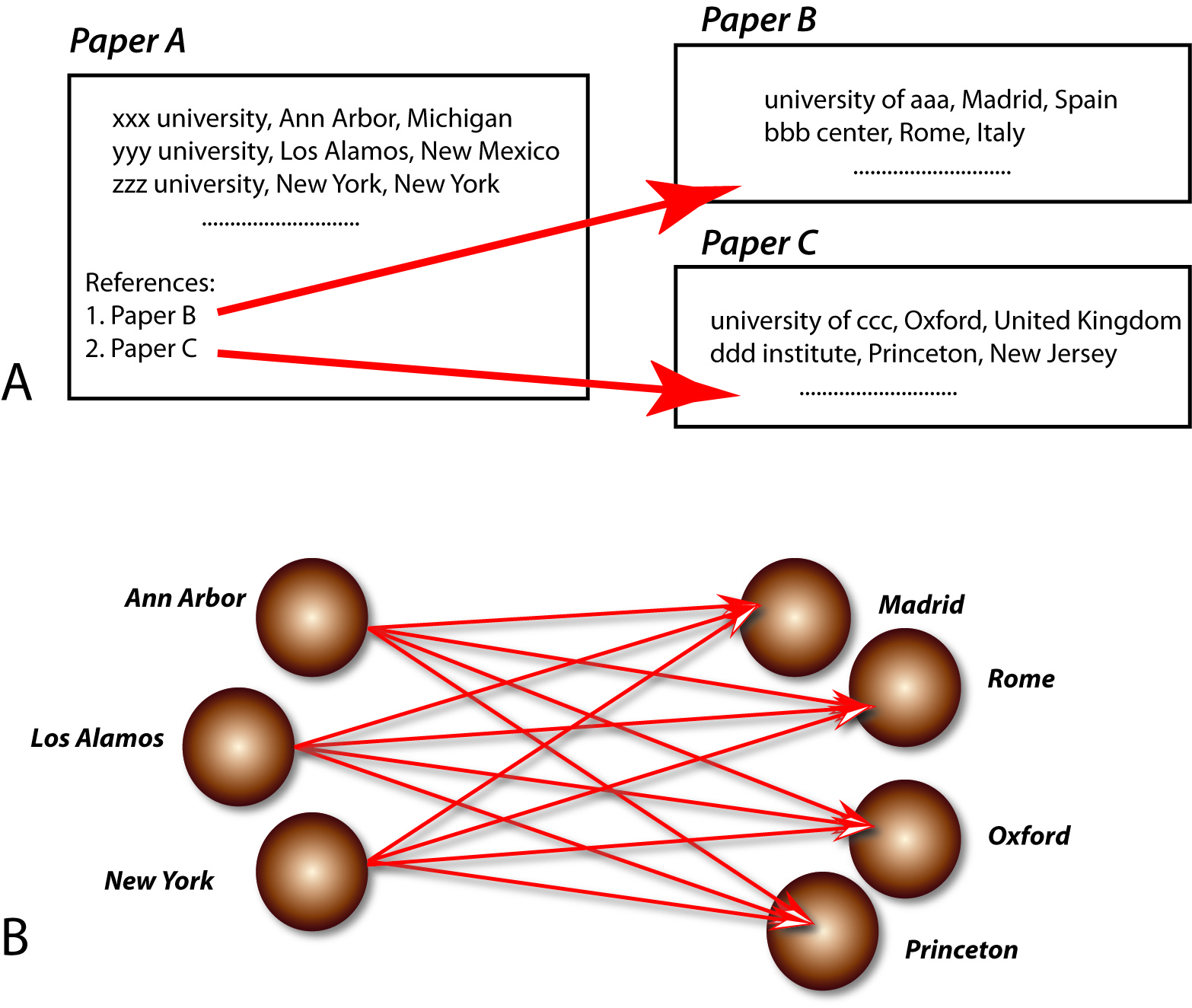}
\caption{{\bf Projecting a paper citation relationship into a city-to-city citation network.} (A) Paper \textit{A} written by authors from Ann Arbor , Los Alamos and New York cites one paper \textit{B} written by authors from Rome and Madrid and another paper \textit{C} from Oxford and Princeton. (B) In a city-to-city citation network, directed links from Ann Arbor to Madrid, Rome, Oxford and Princeton are generated, and similarly Los Alamos and New York are connected to the above four cited cities.}
\label{fig0}
\end{figure}

In order to construct the geolocalized citation network we consider nodes (urban areas) and directed links representing the presence of citations from a paper with affiliation in one urban area to a paper with affiliation in another urban area. For example, if a paper written in node $i$ cites one paper written in node $j$ there is an link from $i$ to $j$, i.e., $j$ receives a citation from $i$ and $i$ sends a citation to $j$. Each paper may have multiple affiliations and therefore citations have to be proportionally distributed between all the nodes of the papers. For this reason we weight each link in order to take into account the presence of multiple affiliations and multiple citations. 
In a given time window, the total number of citations for papers written in $j$ received from papers written in $i$, is the weight of the link $i \rightarrow j$, and the total number of citations for those paper written in $j$ sent to the papers written in $k$ is the weight of the link $j \rightarrow k$. For instance, if in a time window $t$, there is one paper written in node $j$, which cite two papers written in node $k$ and was cited by three papers written in node $i$, then  $w_{jk}=2, w_{ij}=3 $, and we add all such weights for each paper written in that node $j$ and obtain the weights for links. For papers written in multiple cities, say $j_1,j_2$, the weight will be counted equally. The time window we use in this manuscript is one year. We show an example of the network construction in Figure~(\ref{fig0}).\\

\begin{figure}[ht]\centering
\includegraphics[width=0.54\columnwidth,angle=0]{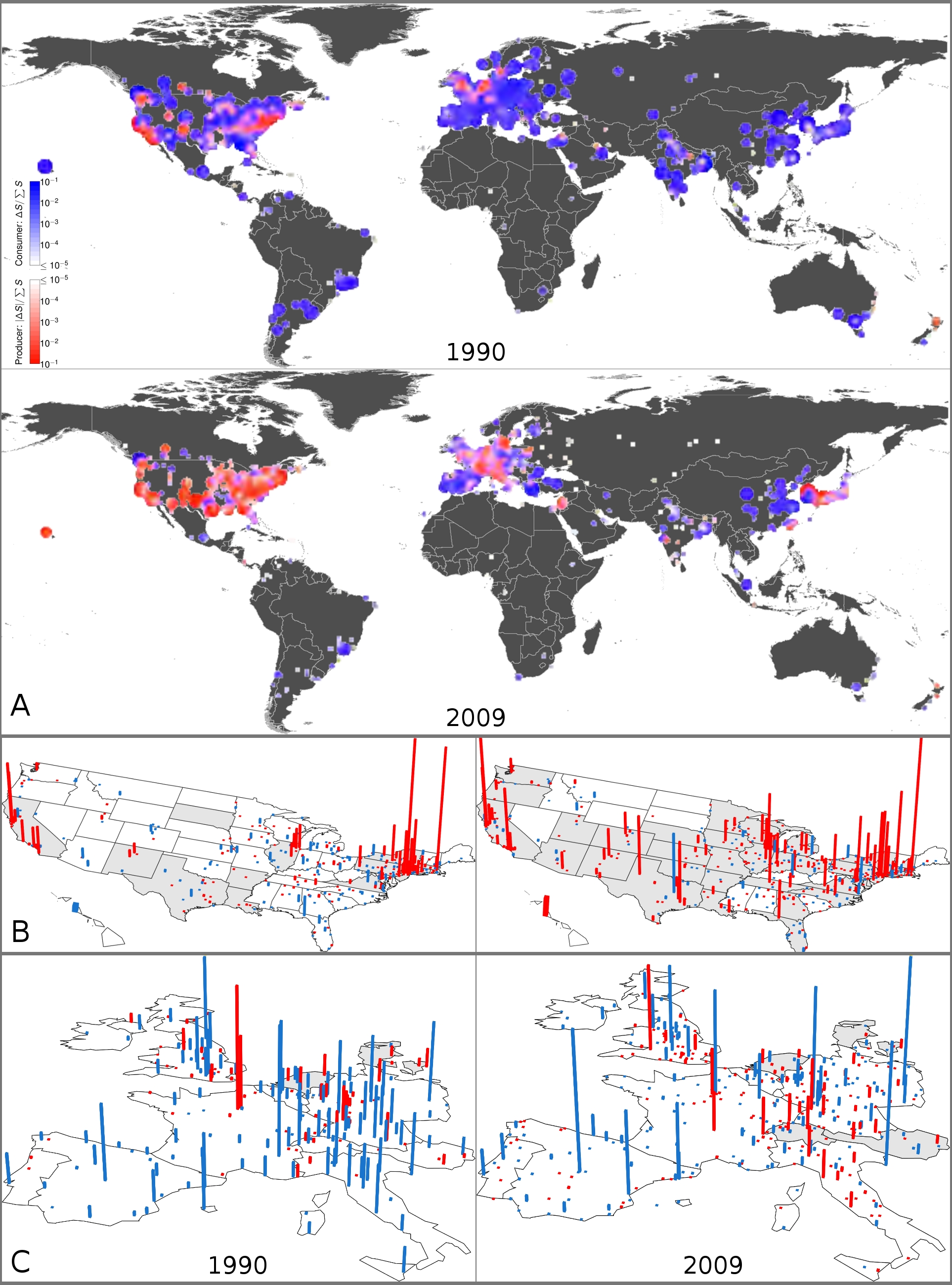}
\caption{{\bf Spatial distributions of scientific producers and consumers of Physics.} The geospatial distribution of scientific producer and consumer cities. (A) The world map of producers and consumers at the city level in $1990$ (top) and $2009$ (bottom). A producer city, of which the relative unbalance $\varDelta S_{i} > 0$, is coloured in red scale.  A consumer with the relative unbalance $\varDelta S_{i} < 0$ is coloured in blue scale. The darkness of colour is proportional to the absolute value of unbalance. The larger the absolute value of unbalance, the darker the colour.  (B) The map of producer and consumer cities in the continental United States in $1990$ (left) and $2009$ (right). (C) The map of producer and consumer cities in selected European countries in $1990$ (left) and $2009$ (right). In (B) and (C), a producer city is marked with a red bar, while a consumer city is marked with a blue bar. The height of each bar is scaled with $|\varDelta S_{i}|$. Note that in (C) the height of bars is not scaled with the height in (B) for visibility. Maps in panel A are created by using ArcGIS\textsuperscript{\textregistered} \cite{arcgis}, and maps in panel B and C are created by using R \cite{R}.}
\label{fig1}
\end{figure}

In order to define main actors in the production and consumption of Physics, we consider citations as a currency of trade. This analogy allows us to immediately grasp the meaning and distinction between producers and consumers of scientific knowledge. Nodes that receive citations export their knowledge to the citing nodes. Instead, nodes that cite, papers produced from other nodes of the network, import knowledge from the cited nodes.  Measuring the unbalance trade between citations, we define \emph{producers} as cities that export more than they import, and \emph{consumers} as cities that import more than they export. More precisely, we can measure the total knowledge imported by each urban area as $\sum_j w_{ij}$ and the total export as $\sum_j w_{ji} $ in a given year. Those measures however acquire specific meaning when considered relatively to the total trade of physics knowledge worldwide in the same year; i.e. the total number of citations worldwide $S=\sum_{ij} w_{ij}$. The relative trade unbalance of each urban area $i$ is then: 
\begin{equation}
\varDelta S_i = \frac{ \sum_j w_{ji}- \sum_j w_{ij}}{S}.
\end{equation}
A negative or positive value of this quantity indicates if the urban area $i$ is consumer or producer, respectively. In Figure~(\ref{fig1})-A we show the worldwide geographical distribution of producer (red) and consumer (blue) urban areas for the   $1990$ and $2009$. Interestingly, during the $90$s the production of Physics knowledge was highly localized in a few cities in the eastern and western coasts of the USA and in a few areas of Great Britain and Northern Europe. In $2009$ the picture is completely different with many producer cities in central and southern parts of the USA,  Europe and Japan. It is interesting to note that despite the fraction of papers produced in the USA is generally decreasing or stable, many more cities in the USA acquire the status of knowledge producers. This implies that the quality of knowledge production from the USA is increasing and thus attracting more citations. This makes it clear that the knowledge produced by an urban area can not be considered to be measured only by the raw number of papers. Citations are a more appropriate proxy that encodes the value of the products. They serve as an approximation of the actual flow of knowledge. The Figure~(\ref{fig1})-A  also makes it clear that cities in China are playing the role of major consumers in both $1990$ and $ 2009$. We also observe that cities in other countries like Russia and India consumed less in $2009$ than $1990$. In other words, in $2009$ both the production and consumption of knowledge are less concentrated on specific places and generally spread more evenly geographically. In order to provide visual support to this conclusion we show in Figure~(\ref{fig1})-B  the geographical distribution of producers and consumers inside the USA. From the two maps it is evident the drift of knowledge production from the two coastal areas in the USA to the midwest, central and southern states. Similarly, in Figure~(\ref{fig1})-C we plot the same information for western Europe.  In $1990$ only a few urban areas in Germany and France were clearly producers. By $2009$ this dominance has been consistently eroded by Italy, Spain and a more widespread geographical distribution of producers in France, Germany and UK.
\subsection*{Knowledge diffusion proxy.}
\begin{figure}[ht]\centering
\includegraphics[width=\columnwidth,angle=0]{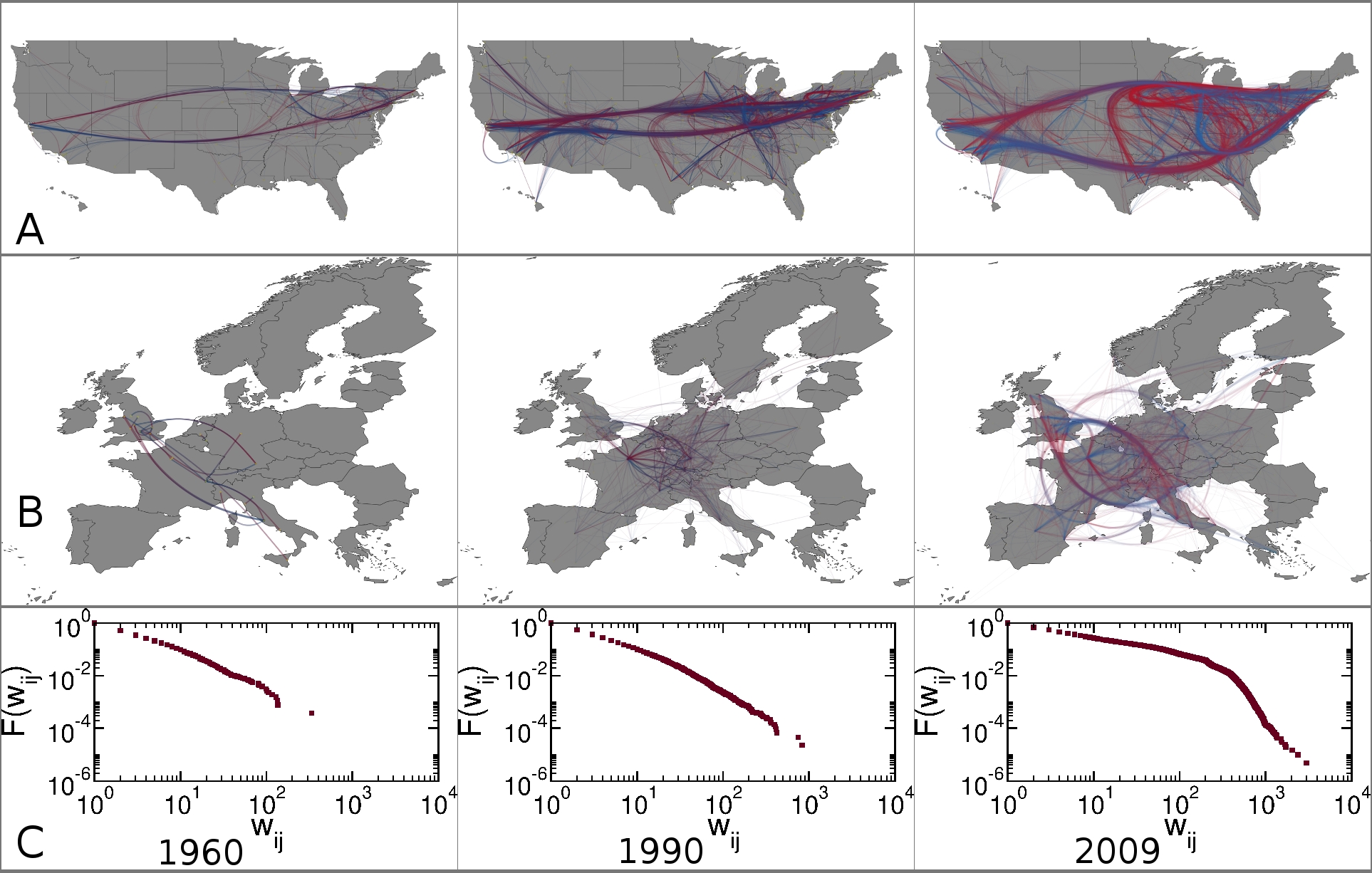}
\caption{{\bf Networks structure.} The network structures of city-to-city citation networks. (A) The backbones ($\alpha=0.1$) of the citation networks at the city level within the United States in $1960$, $1990$, $2009$ (from the left to right). (B) The backbones ($\alpha=1, 0.1, 0.1$ from left to right) of the citation networks at the city level within the European Union 27 countries as well as Switzerland and Norway in $1960$, $1990$, $2009$ (from the left to right). In (A) and (B), the color shows the direction of links: if node $i$ cites node $j$ there is a link starting with blue and ending with red. (C) The cumulative distribution function of the link weights $F_{w}(w_{ij}) = P(w\ge w_{ij})$ for the city-to-city citation networks in year $1960$, $1990$ and $2009$ (from left to right). The maps of networks in (A) and (B) were created using JFlowMap \cite{jflowmap}.}
\label{fig2}
\end{figure}

The definition of producers and consumers is based on a local measure, that does not allow to capture all possible correlations and bounds between nodes that are not directly connected. This might result in a partial view and description of the system, especially when connectivity patterns are complex~\cite{barabasi99-3,barrat08-1,newman10-1,alex09-1,alex12-1}. Interestingly, a close analysis of each citation network, see Figure~(\ref{fig2}), clearly shows that citation patterns have indeed all the hallmarks of complex systems~\cite{barabasi99-3,barrat08-1,newman10-1,alex09-1,alex12-1}, especially in the last two decades. The system is self-organized, there is not a central authority that assigns citations and papers to cities, there is not a blueprint of system's interactions, and as clearly shown from Figure~(\ref{fig2})-C the statistical characteristics of the system are described by heavy-tailed distributions~\cite{barabasi99-3,barrat08-1,newman10-1,alex09-1,alex12-1}. Not surprisingly, the level of complexity of the system has increased with time. In Figure~(\ref{fig2})-A we plot the most statistically significant connections of the citation network between cities inside USA in $1960$, $1990$ and $2009$. We filter links by using the backbone extraction algorithm~\cite{backbone} which preserves the relevant connections of weighted networks while removing the least statistically significant ones.  We visualize each filtered network by using a bundled representation of links~\cite{jflowmap}. The direction of each weighted link goes from blue (citing) to red (cited). Similarly, in Figure~(\ref{fig2})-B, we visualize the most significant links between cities in Europe (European Union's 27 countries, as well as Switzerland and Norway). It is clear from Figure~(\ref{fig2})-A that in $1960$ the citation patterns inside the USA were limited to a few cities, and in Europe only a few cities were connected. Instead, in $1990$ and $2009$ we register an increase in the interactions among a larger number of cities. The observed temporal trend is well known and valid not just for Physics~\cite{Adams1996}.  Among many factors that have been advocated to explain this tendency we find the increase of the research system and the advance in technology that make collaboration and publishing easier~\cite{Rosenblat2004,Havemann2006,Agrawal2008,Adams2012}. \\

In order to explicitly consider the complex flow of citations between producers and consumers, we propose the knowledge diffusion proxy algorithm (see Methods section for the formal definition). In this algorithm, producers inject citations in the system that flow along the edges of the network to finally reach consumer cities where the injected citations are finally absorbed. The algorithm allows charting the diffusion of knowledge, going beyond local measures. The entire topology of the networks is explored uncovering nontrivial correlations induced by global citation patterns. For instance, knowledge produced in a city may be consumed by another producer that in turn produces knowledge for other cities who are consumers. This points out that the actual consumer of knowledge is not just signalled by the unbalance of citations but in the overall topology of the production and consumption of knowledge in the whole network.  Indeed, the final consumer of each injected citation may not be directly connected with the producer. Citations flow along all possible paths, sometimes through intermediate cities. In Table~(\ref{table1}), and Table~(\ref{table2}) we report the rankings of Top $10$ final consumers evaluated by the knowledge diffusion proxy for the Top $3$ producers in $2009$ and $1990$ respectively. We also list the Top $10$ neighbours according to the local citation unbalance.  From these two tables, it is clear that the final rank of each consumer, obtained by our algorithm, can be extremely different from the ranking obtained by just considering local unbalances. For instance, in $2009$ Bratislava and Mainz rank in top $10$ consumers absorbing knowledge produced in Boston. However, according to local measure of unbalance, these two cities are ranked out of top $10$ (shown in bold in Table~(\ref{table1})).
%For instance, in $2009$ knowledge produced in Boston finally flows into some cities that are not directly connected with it such as Bratislava, Mainz, etc. (shown in bold in Table~(\ref{table1})) through a chain of citations. 
Interestingly, even the Top consumer for New Haven, Berlin, also does not rank among the Top $10$ neighbours according to the citation unbalance. These findings confirm that in order to uncover the complex set of relationships among cities, it is crucial to consider the entire structure of the network, going beyond simple local measures. \\
\begin{table}[ht]
\centering
\caption{Rankings from Knowledge diffusion proxy algorithm for top 3 producer cities in 2009. In bold, we highlight cities that are present in top 10 consumers ranked according to the knowledge diffusion proxy but do not appear in top 10 cities ranked according to local citation unbalance. }
\scalebox{0.75}{
\begin{tabular}{llllll}
%\begin{tabular*}{1\hsize}{@{\extracolsep{\fill}}*{6}{l}}
\hline 
\multicolumn{2}{c}{Boston} & \multicolumn{2}{c}{Berkeley} & \multicolumn{2}{c}{New Haven} \\ 
Diffusion proxy & Citation unbalance & Diffusion proxy & Citation unbalance & Diffusion proxy & Citation unbalance\\ 
 \hline 
Athens & Madrid & Athens  & Athens &\textbf{Berlin} & Vancouver \\ 
Madrid & Athens & Gwangju & Madrid & Athens & Paris \\ 
Vancouver & Vancouver & Bratislava  & Bratislava & \textbf{Mainz} & Trieste \\ 
Gwangju & Moscow & Madrid & Paris & Vancouver & Athens \\ 
\textbf{Bratislava}  & Paris & Vancouver & Vancouver & Gwangju & Gwangju \\ 
Berlin  & Tokyo & Trieste & Gwangju & Trieste & Bratislava  \\ 
Trieste & Trieste & Waco & Moscow  & Bratislava  & Madrid \\ 
\textbf{Mainz} & Beijing & Paris & Trieste & \textbf{Coventry} & Liverpool \\ 
Paris & Berlin & \textbf{Berlin} & Seoul & \textbf{Valencia} & Oxford \\ 
\textbf{Waco} & Gwangju & \textbf{Mainz} & Waco & \textbf{Madrid} & Santa Barbara \\ 
\hline 
%\end{tabular*}
\end{tabular}}
\label{table1}
\end{table}

\begin{table}[ht]
\centering
\caption{Rankings from Knowledge diffusion proxy algorithm for top 3 producer cities in 1990. In bold, we highlight cities that are present in top 10 consumers ranked according to the knowledge diffusion proxy but do not appear in top 10 cities ranked according to local citation unbalance. }
\scalebox{0.75}{
\begin{tabular}{llllll}
\hline 
\multicolumn{2}{c}{Piscataway} & \multicolumn{2}{c}{Boston} & \multicolumn{2}{c}{Palo Alto} \\ 
Diffusion proxy & Citation unbalance & Diffusion proxy & Citation unbalance & Diffusion proxy & Citation unbalance\\ 
 \hline 
Tokyo & Stuttgart & Tokyo & Tokyo	& Tokyo & Tokyo \\ 
\textbf{Beijing} & Tokyo & Grenoble & Grenoble 	& \textbf{Beijing} & Ann Arbor \\ 
\textbf{Tsukuba} & Los Angeles & \textbf{Beijing} & Los Angeles & \textbf{Tsukuba} & Bloomington \\ 
Grenoble & Urbana & \textbf{Tsukuba} & College Park & Seoul & Boulder \\ 
\textbf{Tallahassee} & College Park & \textbf{Seoul} & Los Alamos & \textbf{Tallahassee} & Urbana \\ 
Hamilton  & Grenoble & Vancouver & Urbana & \textbf{Charlottesville} & Berlin \\ 
\textbf{Buffalo} & Rochester  & \textbf{Tallahassee} & Boulder & \textbf{Vancouver} & Orsay \\ 
\textbf{Vancouver} & Boston & \textbf{Warsaw} & Rochester & Berlin  & Denver \\ 
\textbf{Charlottesville} & Los Alamos & \textbf{Kolkata} & Vancouver 	& \textbf{Durham} & Seoul \\ 
\textbf{Tempe} & Hamilton & \textbf{Charlottesville} & Bloomington & \textbf{Taipei} & Los Alamos \\ 
\hline 
\end{tabular}
}
\label{table2}
\end{table}

\begin{figure}[ht]\centering
\includegraphics[width=0.54\columnwidth,angle=0]{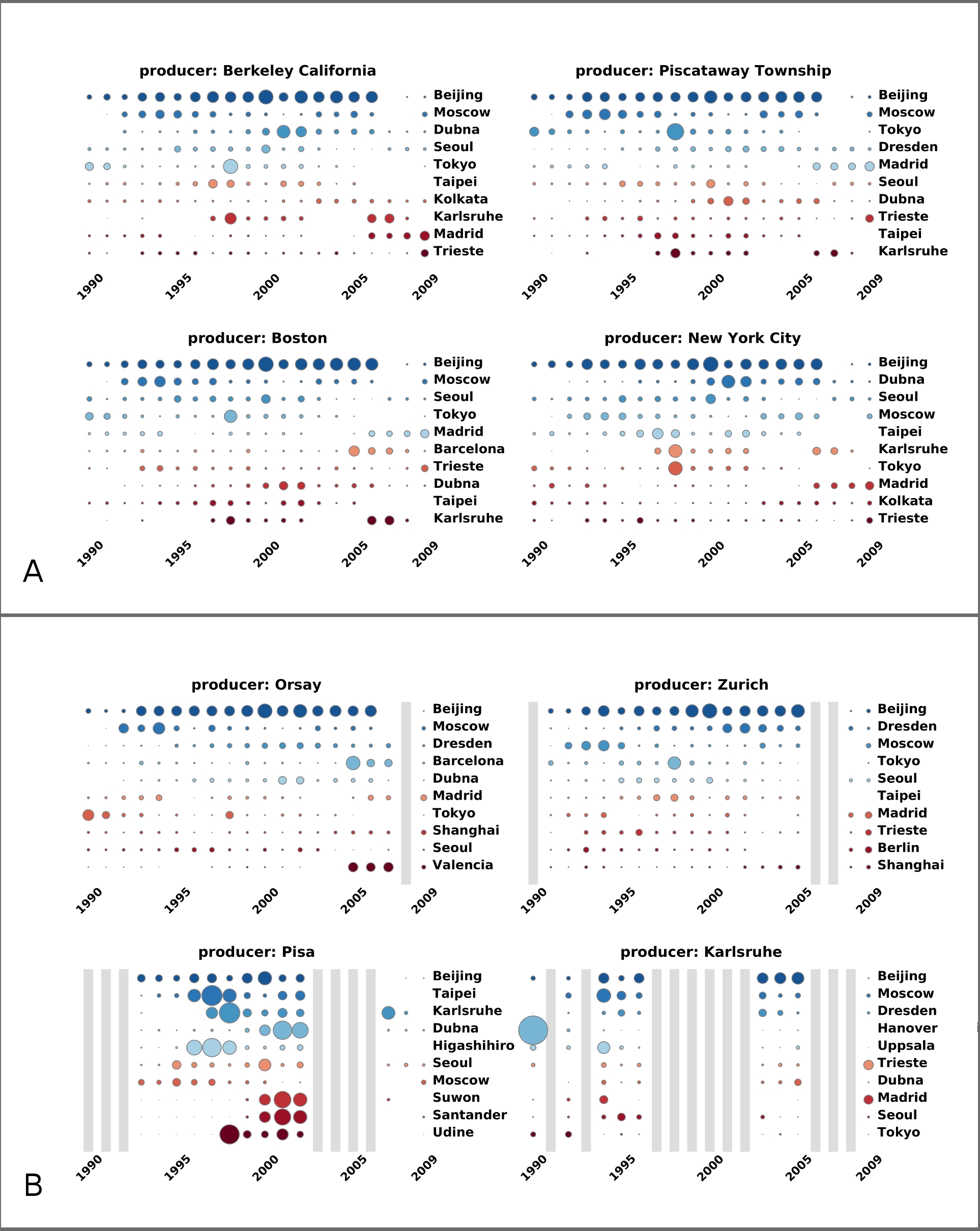}
\caption{{\bf Knowledge diffusion proxy results.} (A) The Top $4$ producer cities in the USA in $2009$ and their Top $10$ consumers from knowledge diffusion proxy algorithm in $1990-2009$. (B) The Top $4$ producer cities in the European Union 27 countries as well as Switzerland and Norway in $2009$ and their Top $10$ consumers from knowledge diffusion proxy algorithm in $1990-2009$. When a producer city becomes a consumer in some year, a grey strip is marked in that year. For each producer city in (A) and (B), the major consumers of the first producer city $m$ in $20$ years are plotted as a function of time from $1990$ to $2009$. The size of the bubble in position $(Y, c)$ is also proportional to the counter $g_{m,c}(Y)$ in that year. The consumer cities for each producer are ordered according to the total number 
of counters in 20 years, i.e., $\sum_{Y_{\mathrm{min}}}^{Y_{\mathrm{max}}}g_{m,c}(Y)$. }
\label{fig3}
\end{figure}

In Figure~(\ref{fig3})-A and Figure~(\ref{fig3})-B we visualize the results considering the Top four producer cities in $2009$ in the USA and in Europe respectively. We show their Top ten consumers over $20$ years as function of time. The size of each circle is proportional to how many times each injected citation is absorbed by that consumer. In the plot, vertical grey strips indicate that the city was not a producer during those years (e.g. Orsay in $2008$). The results show that, on average, Beijing is the top consumer for all of these producers in the past $20$ years. Since China registered a big economical growth and increment of research population in the early $2000$, it is reasonable to assume that, thanks to this positive stimulus, many more papers were written in its capital, a dominant city for scientific research in China. However, the fast publication growth increased the unbalance between sent and received citations. Each paper published in a given city imports knowledge from the cited cities. Reaching a balance might require some time. Each city needs to accumulate citations back to export its knowledge to others cities. We can speculate that in the near future cities in China might be moving among the strongest producers if a fair number of papers start receiving enough citations, which obviously depends on the quality of the research carried out in the last years. This is the case of cities like Tokyo which has gradually approached the citation balance in recent years. For instance, Table~(\ref{table2}) shows that in $1990$ Tokyo, was among the top consumers. But by $2009$, its contribution to citation consumption had become less significant as observed from Figure~(\ref{fig3}) and Table~(\ref{table1}).
% \nicola{Probably confusing} We also observe that as country China is the largest consumer for the top four producer countries if we perform the same experiment on country-to-country network (see SI). The comparison between two different granularities shows significant correlations between the behavior of a country and its major cities. However, the temporal patterns are sometime shifted. In the case of China for example, the transition from being consumer to becoming, eventually, a producer started in Beijing much earlier. The capital in this case pushed the production of the country more than other cities.
\subsection*{Ranking Cities.}

\begin{figure}[ht]\centering
\includegraphics[width=\columnwidth,angle=0]{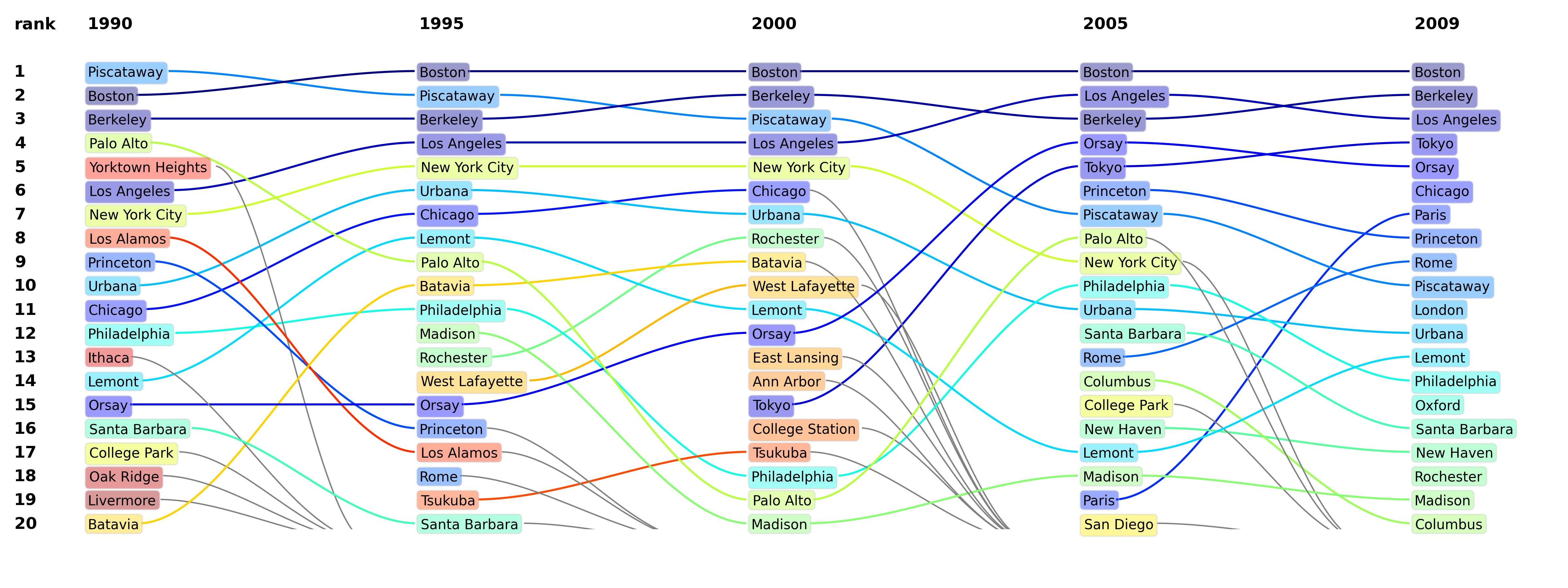}
\caption{{\bf Top 20 ranked cities as a function of time.} The plot summarizes Top $20$ ranked cities in $1990$, $1995$, $2000$, $2005$ and $2009$ (from left to right), and relations between the rankings in different years. The grey lines are used when the rank of that city drops out of Top $20$.}
\label{fig4}
\end{figure}

\begin{figure}[ht]\centering
\includegraphics[width=\columnwidth,angle=0]{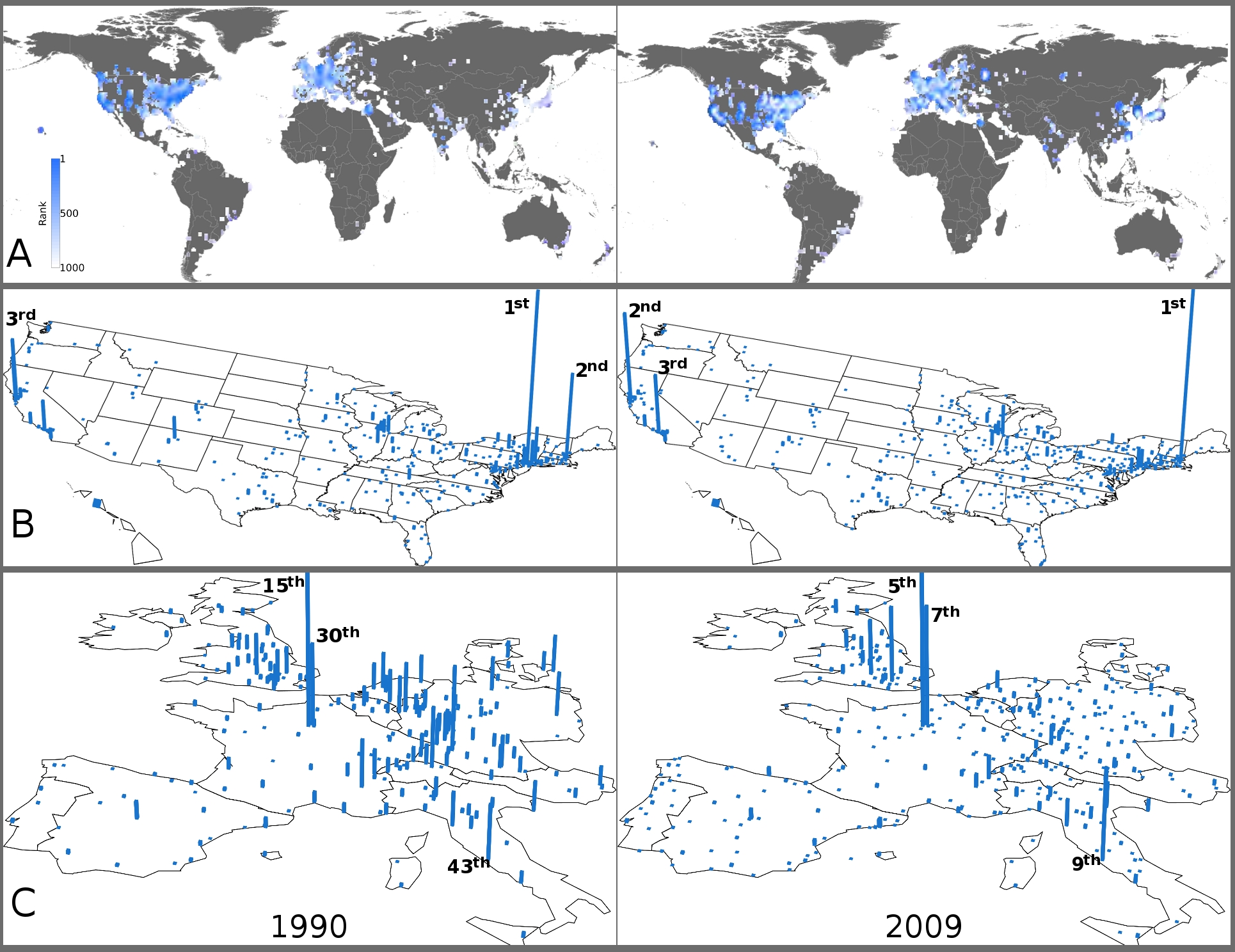}
\caption{{\bf Geospatial distribution of city ranks.} (A) The world map of city ranks in $1990$ (left) and $2009$ (right). The ranking of each city is represented by color from blue (high ranks) to white (low ranks). (B) The map of ranks for cities in the United States in $1990$ (left) and $2009$ (right). (C) The map of ranks for cities in the selected European countries in $1990$ (left) and $2009$ (right). In (B) and (C), each city is marked with a bar, and the height of each bar is inversely proportional to the ranking position. The Top $3$ rank positions in each region are labelled for reference. Note that in (C) the height of bars is not scaled with the height in (B) for visibility. Maps in panel A are created by using ArcGIS\textsuperscript{\textregistered} \cite{arcgis}, and maps in panel B and C are created by using R \cite{R}.}
\label{fig5}
\end{figure}

\begin{figure}[ht]\centering
\includegraphics[width=0.54\columnwidth,angle=0]{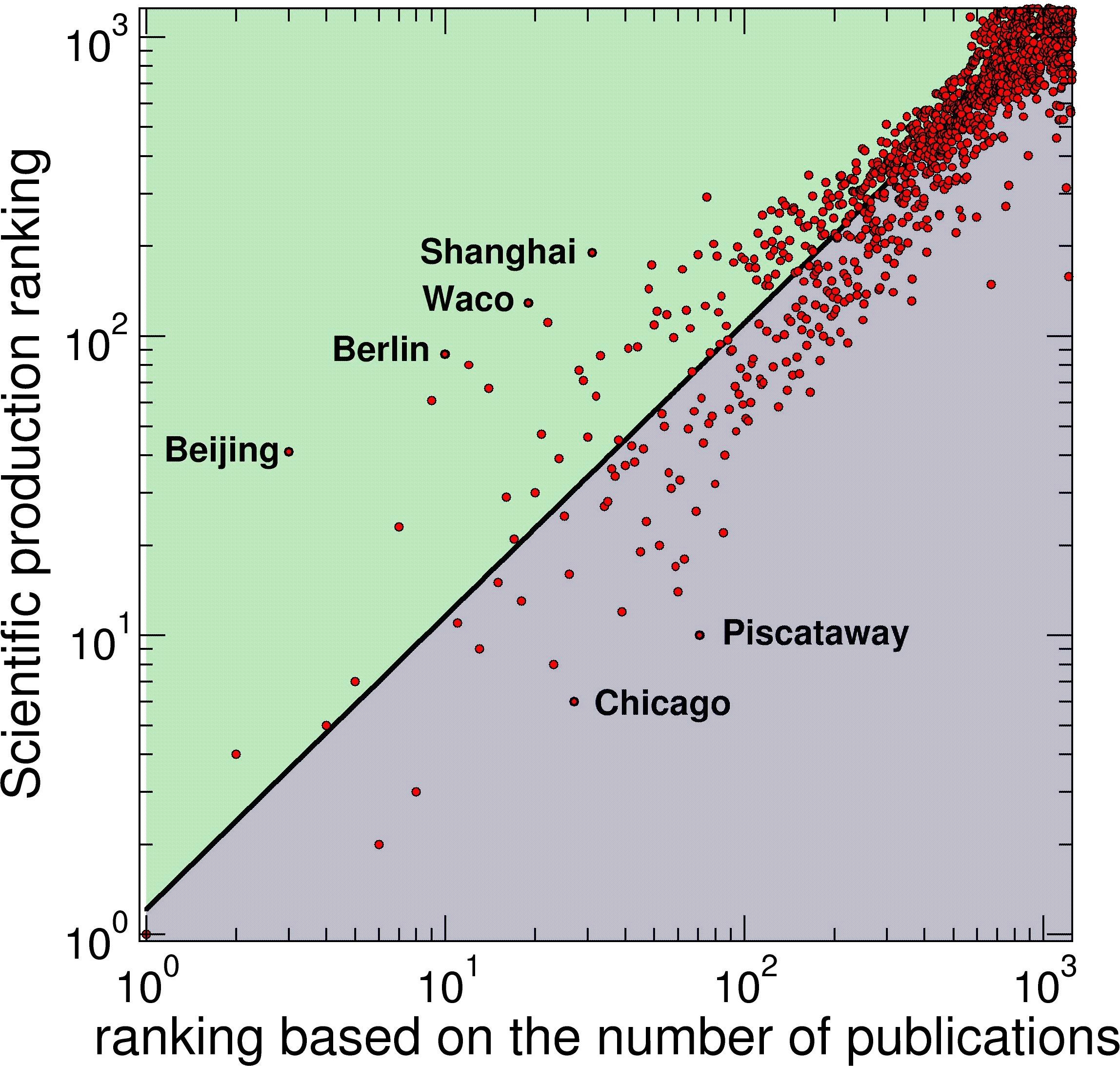}
\caption{{\bf Correlation between scientific production ranking and ranking based on the number of publications in $2009$.} The x-axis represents rankings based on the number of papers each city published in $2009$, and the y-axis represents the scientific production ranking for each city in $2009$. The solid line corresponds to the power-law fitting of data with slope $-0.98$, and separates the space into two regions. In the region below the line (coloured blue), cities gain better rankings from scientific production ranking algorithm even with relatively less publications, such as Chicago and Piscataway. In the region above (coloured green) cities have lower rankings from the algorithm even they have more papers published, such as Beijing, Berlin, Wako and Shanghai. }
\label{fig6}
\end{figure}

Authors, departments, institutions, government and many funding agencies are extremely interested in defining the most important sources of knowledge. The necessity to find objective measures of the importance of papers, authors, journals, and disciplines leads to the definition of a wide variety of rankings~\cite{Garfield1979,Egghe1990}. Measures such as impact factor, number of citations and h-index~\cite{Hirsch2005} are commonly used to assess the importance of scientific production. However, these common indicators might fail to account for the actual importance and prestige associated to each publication. In order to overcome these limitations, many different measures have been proposed~\cite{Kleinberg1999,CiteRank,Castillo2007,Sidiropoulos2007}. Here we introduce the {\em scientific production ranking algorithm} (SPR), an iterative algorithm based on the notion of diffusing scientific credits. It is analogous to  PageRank~\cite{brin98-1}, CiteRank~\cite{CiteRank}, HITS~\cite{Kleinberg1999}, SARA~\cite{SARA}, and others ranking metrics. In the algorithm each node receives a credit that is redistributed to its neighbours at the next iteration until the process converges in a stationary distribution of credit to all nodes (see Methods section for the formal definition). The credits diffuse following citations links self-consistently, implying that not all  links have the same importance. Any city in the network will be more prominent in rank if it receives citations from high-rank sources. This process ensures that the rank of each city is self-consistently determined not just by the raw number of citations but also if the citations come from highly ranked cities.  In Figure~(\ref{fig4}) we show the Top $20$ cities from $1990$ to $2009$. Interestingly, we clearly see the decline and rise of cities along the years as well as the steady leadership  of Boston and Berkeley. This behaviour is clear in Figure~(\ref{fig5})-B where we show the rank for cities in USA in $1990$ and $2009$. Meanwhile, the ranking of cities in European and Asian countries like France, Italy and Japan has increased significantly, as shown in both Figure~(\ref{fig4}) and Figure~(\ref{fig5})-A. In Figure~(\ref{fig5})-C we focus on the geographical distribution of ranks for a selected set of European countries in $1990$ and $2009$. In Table~(\ref{table3}) we provide a quantitative measure of the change in the landscape of the most highly ranked cities in the world by showing the percentage of cities in the top 100 ranks for different continents. In Figure~(\ref{fig6}), we compare the ranking obtained by our recursive algorithm with the ranking obtained by considering the total volume of publications produced in each city. Since we are considering only journals by the APS, the impact factor is consistent across all cities and does not include disproportionate effects that often happen when mixing disciplines or journal with varied readership. It is then natural to consider a ranking based on the raw productivity of each place. As we see in the figure though the two rankings, although obviously correlated, provide different results. A number of cities whose ranking, according to productivity, is in the Top 20 cities in the world, are ranked one order of magnitude lower by the SPR algorithm. Valuing the number of citations and their origin in the ranking of cities produces results often not consistent with the raw number of papers, signaling that in some places a large fraction of papers are not producing knowledge as they are not cited. We believe that the present algorithm may be considered as an appropriate way to rank scientific production taking properly into account the impact of papers as measured by citations. 

\begin{table}[ht]
\centering
\caption{Percentage of top 100 ranked cities in continents in 1990 and 2009.}
\begin{tabular*}{0.45\textwidth}{@{\extracolsep{\fill} }lrr}
\hline
Continent   & 1990    &  2009 \\
\hline
Asia        & 4.0\%    & 11.0\% \\
Europe      & 24.0\% & 33.0\% \\
N. America  & 72.0\% & 56.0\% \\
\hline
\end{tabular*}
\label{table3}
\end{table}

\section*{Discussion}
In this paper we study the scientific knowledge flows among cities as measured by papers and citations contained in APS~\cite{aps10-1} journals. In order to make clear the meaning and difference between producers and consumers in the context of knowledge, we propose an economical analogy referring to citations as a traded currency between urban areas. We then study the flow of citations from producers to consumers with the knowledge production proxy algorithm. Finally, we rank the importance of cities as function of time using the scientific production ranking algorithm. This method, inspired by the PageRank~\cite{brin98-1}, allows us to evaluate the importance of cities explicitly considering the complex nature of citation patterns. In our analysis we considered just scientific publications contained in the APS journals~\cite{aps10-1}.  We do not have information on citations received or assigned to papers outside this dataset. These limitations certainly affect the count of citations of each city, potentially creating biases in our results. However, our findings, while limited to a particular dataset, are aligned with different observations reported by other studies focused on other datasets and fields. For example, we identify major US cities (e.g. Boston and San Francisco areas),  as the most important sources of Physics. Similar observations have been done by B\"{o}rner et al.~\cite{Borner2006} at the institution level considering papers published in the Proceedings of the National Academy of Sciences, by Mazloumian et al.~\cite{Boner2013} at country and city level with Web of Science dataset, and by Batty~\cite{Batty03} at both institution and country level considering the Institute for Scientific Information (ISI) \textit{HighlyCited} database. We also find that some European, Russian and Japanese cities have gradually improved their productivities and ranks in recent twenty years. Similarly, such growth in scientific production has been observed by King~\cite{King2004} in the ISI database. As discussed in detail in the SI, by aggregating citations of cities to their respective countries, we find the same correlation between the number of citations, as well as the number of papers, and the GDP invested on Research and Development of several countries as reported by Pan et al.~\cite{Fortunato2012} based on the ISI database. This analogy between our results, and many others in the literature, suggests that the APS dataset, although limited, is representative of the overall scientific production of the largest countries and cities in the recent 20 years. 
The methodology proposed in this paper could be readily extended to larger datasets for which the geolocalization of multiple affiliation is possible. In view of the different rate of publications and citations in different scientific fields we believe however that the analysis of scientific knowledge production should only consider homogeneous datasets. This would help the understanding of knowledge flows in different areas and identify the hot spot of each discipline worldwide. 

\section*{Methods}
\subsection*{Dataset.}

The dataset of the American Physical Society journals, considering papers published between $1893$ and $2009$ of which $450,655$ papers include a list of affiliations~\cite{aps10-1}. Each of paper may have multiple affiliations. In total there are $945,767$ affiliation strings.\\

In order to geolocalize the articles, we parse the city names from the affiliation strings for each article. First, we process each affiliation string and try to match country or US state names from a list of known names and their variations in different languages. We crosscheck the results with Google Map API obtaining validated location information for $97.7\%$ of affiliation strings, corresponding to $445,223$ articles. It is worth noticing that we do not use Google Map API (or other map APIs like Yahoo! or Bing) directly for geocoding because, to our best knowledge, there are no accuracy guarantees to these API results. For each affiliation string with an extracted country or state name, we also match the city name against GeoName database \cite{GeoName} corresponding to its country or US state. $92.6\%$ of affiliation strings with extracted city names are subsequently verified with Google Map API. Finally, a total of $425,233$ publication articles successfully pass the filters we describe here.\\

The dataset also provides $4,710,548$ records of citations between articles published in APS journals. To build citation networks at the city level, we merge the citation links from the same source node to the same target node, and put the total citations on this link as the weight. For articles with multiple city names, the weight will be equally distributed to the links of these nodes. There are totally $2,765,565$ links for city-to-city citation networks from $1960$ to $2009$. (For the full details of parsing country and city names, as well as building networks, see Supplementary Information (SI))

\subsection*{Knowledge diffusion proxy algorithm.}

This analysis tool is inspired by the \emph{dollar experiment}, originally developed to characterized the flow of money in economic networks~\cite{DollarExperiment}. Formally, it is a biased random walk with sources and sinks where a citation diffuses in the network. The diffusion takes place on top of the network of net trade flows. Let us define $w_{ij}$ as the number of citation that node $i$ gives to $j$ and $w_{ji}$ as the opposite flow. We can define the antisymmetric matrix $T_{ij}=w_{ij}-w_{ji}$. The network of the net trade is defined by the matrix $\mathbf{F}$ with $F_{ij}=|T_{ij}|=|T_{ji}|$ for all connected pairs $(i,j)$ with $T_{ij}<0$ and $F_{ij}=0$ for all connected pairs $(i,j)$ with $T_{ij}\ge0$. There are two types of nodes. Producers are nodes with a positive trade unbalance $\varDelta s_i=s_i^{in}-s_i^{out}=\sum_jF_{ji}-\sum_jF_{ij}$. Their strength-in is larger than their strength-out. On the other hand, consumers are nodes with a negative unbalance $\varDelta s$. On top of this network a citation is injected in a producer city. The citation follows the outgoing edges with a probability proportional to their intensities, and the probability that the citation is absorbed in a consumer city $j$ equals to $P_{\mathrm{abs}}(j)=\varDelta s_j/s^{in}_j$. By repeating many times this process from each starting point (producers) we can build a matrix with elements $e_{ij}$ that measure how many times a citation injected in the producer city $i$ is absorbed in a city consumer $j$.

\subsection*{Scientific production ranking algorithm.}
The scientific production rank is defined for each node $i$
according to this self-consistent equation:
\begin{equation}
P_i =  q
z_i+\left(1-q\right)\sum_j \; \frac{P_j}{s^{out}_j} w_{ji} + \left(1-q\right) z_i \sum_j \; P_j \; \delta \left( s^{out}_j
\right).
\label{eq:pg}
\end{equation}
$P_i$ is the score of the node $i$, $0 \leq q \leq 1$
is the damping factor (defining the probability of random jumps reaching any other node in the network), $w_{ji}$ is the weight of the directed
connection from $j$ to
$i$, $s^{out}_j$ is the strength-out of the node $j$ and finally $\delta (x) ,$ is the Dirac delta function that is $0$ for $x=0$ and $1$ for $x=1$.
 Here we use the damping factor $q=0.15$. The first term on the r.h.s. of Eq.~(\ref{eq:pg}) defines the redistribution of credits to all nodes in the network due to the random jumps in the diffusion. The second term defines the diffusion of credit through the network. Each node $i$ will get a fraction of credit from each citing node $j$ proportional to the ratio of the weight of link $j\to i$ and the strength-out of node $j$. Finally the last term defines the redistribution of credits to all the nodes in the networks due to the nodes with zero strength-out. In the original PageRank the vector $\mathbf{z}$ has all the components equal to $1/N$ (where $N$ is the total number of nodes). Each component has the same value because the jumps are homogeneous. In this case instead, the vector $\mathbf{z}$ considers the normalized scientific credit given
to the node $i$ based on his productivity. Mathematically we have:
\begin{equation}
z_i = \frac{\sum_p \delta_{p,i} \; 1/n_p}{\sum_j \sum_p \delta_{p,j} \; 1/n_p} \;\;\; ,
\label{eq:credit}
\end{equation}
where $p$ defines the generic paper and $n_p$ the number of nodes who have written the paper. It is important to notice that
$\delta_{p,i}=1$ only if the $i$-th node wrote the paper $p$, otherwise it equals zero.\\
\newline

{\bf Acknowledgments}\\
This work has been partially funded by NSF CCF-1101743 and NSF
CMMI-1125095 awards. We acknowledge the American Physical Society for providing the data about Physical Review's journals.\\
{\bf Author Contributions } \\
A.V., N.P. \& Q.Z. designed research, Q.Z., B.G., \& F.C. parsed data, Q.Z., N.P. \& A.V. analysed data. All authors wrote, reviewed and approved the manuscript.\\
{\bf Competing financial interests}\\
The authors declare no competing financial interests.

\clearpage
\begin{appendices}
%\appendix
\renewcommand{\appendixname}{Supplementary Information}
\renewcommand{\thesection}{\arabic{section}}
\section{Extracting Geographic Information}
The database of Physical Review publications used in this paper consists of $463,348$ articles, each of which is identified by a unique Digital Object Identifier (DOI). $83\%$ of these articles ($450,655$) record the publishing year, the author(s) of the article, as well as the corresponding affiliation(s). An article may have more than one affiliation, and the database provides affiliation strings for each article. In total, we have $945,767$ affiliation strings, and we aim to extract country and city information from the affiliation strings for each article.\\

\begin{figure}[ht]
\centering
 \includegraphics[width=0.55\textwidth]{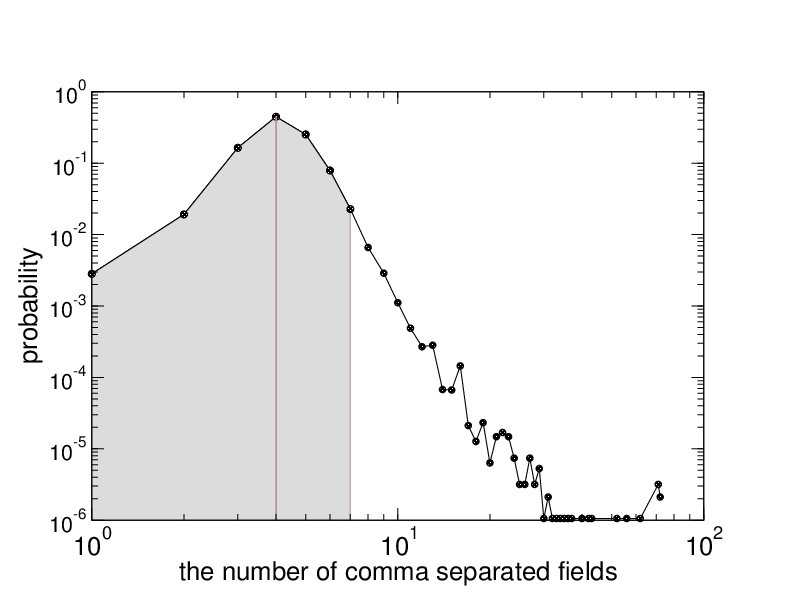}
 \caption{The probability distribution of the number of comma separated fields in an affiliation string. The mean value of such the number is $4.33$ and the standard deviation is $1.156$. The grey area in the plot represents the band with the width of 3 standard deviations, which implies that the most of affiliation strings consist of no more than 7 comma separated fields.}
 \label{fig:appendix:fieldDistribution}
\end{figure}

We observe that an affiliation string likely stands for a single affiliation, roughly consisting of several comma separated fields:
\begin{itemize}
 \item[]
\begin{footnotesize}
\texttt{(SUB-INSTITUTE)*, (INSTITUTE), (OTHER INFORMATION)*, (CITY), (OTHER INFORMATION)*, (COUNTRY/STATE)}
\end{footnotesize}
\end{itemize}

where `\texttt{SUB-INSTITUTE}' means department, college, institute, laboratory within an institute, the asterisk refers to any repetition of the field (including zero), and `\texttt{OTHER INFORMATION}' usually means the province (or region) name, postal codes, or P. O. Box. For instance,
\begin{itemize}
 \item[] \texttt{PHYSICS DEPARTMENT, THE ROCKEFELLER UNIVERSITY, NEW YORK, NEW YORK}
 \item[] \texttt{THE INSTITUTE FOR PHYSICAL SCIENCES, THE UNIVERSITY OF TEXAS AT DALLAS, P. O.BOX 688, RICHARDSON, TEXAS}
 \item[] \texttt{PHYSICS DEPARTMENT, UNIVERSITY OF GUELPH, GUELPH, ONTARIO N1G 2W1, CANADA}
\end{itemize}
\autoref{fig:appendix:fieldDistribution} shows the probability distribution of the number of comma separated fields for all affiliation strings. The mean value of such numbers is $4.33$ and the standard deviation is $1.156$. $86\%$ of all affiliation strings have between 3 and 5 comma separated fields, while the percentage rises to $97\%$ for those with less than 8 such fields (mean$\pm 3\sigma$). Therefore, we first assume that an affiliation string with no more than 7 comma separated fields represents a single affiliation, and the remaining ones may consist of multiple affiliations.

\subsection{Parsing country names}
We first extract country and U.S. state names from single affiliation strings. To find country names, we create a dataset of country names except U.S. from ISO 3166 country codes \cite{ISO3166}, and the name of U.S. states from Wikipedia \cite{USstatesWiki}. For some historical country names in the 20th century (e.g., the Soviet Union, Yugoslavia, East Germany), we manually add them in the dataset. Besides, for some countries, we take into consideration the name variations, like full official names and the name in its official language, and possible abbreviations, e.g., U.S.S.R for the Soviet Union, People's Republic of China for China, Deutschland for Germany, etc.\\

Based on the above assumptions and observations, for an affiliation string with no more than 7 comma separated fields, we first search the field representing a country name, the process of which is called `\textit{field match}'. For each field in an affiliation string, we eliminate the words with numbers 0-9, which may represent a postal code, and then try to match the field with any of the country name in our country name dataset.\\

If there is no field match for an affiliation string, it is possible that either the author did not write a country name specifically but some other fields, like the institution name, include a country name (e.g., \texttt{RANDAL MORGAN LABORATORY OF PHYSICS, UNIVERSITY OF PENNSYLVANIA}), or the country name is mixed with other information in a field, like a city name or a non-numeric postal code (e.g., \texttt{MAX-PLANCK-INSTITUT F\"{U}R MOLEKULARE PHYSIOLOGIE POSTFACH 500247 D-44202 DORTMUND GERMANY}). Moreover, for the affiliation strings with `\textit{field match}' results, other fields in that string may also contain country names for multiple affiliation cases (e.g., \texttt{ARGONNE NATIONAL LABORATORY, ARGONNE, ILLINOIS 60439 AND OHIO STATE UNIVERSITY, COLUMBUS, OHIO}). For the kind of affiliation strings without field match results, we try to match the country name word by word in all fields in that affiliation strings, and for the ones with some field matched, we match the country names word by word in other fields. We call this process `\textit{string match}'. If there is a single match from the above two steps, we assign the matched country name to this affiliation string, and classify it into affiliation strings with unique country name. If there are multiple country names matched, we set these affiliation strings aside for later processing.\\

The above two procedures of `\textit{field match}' and `\textit{string match}' give unique country name to $95.11\%$ affiliation strings ($899,575$ out of $945,767$), but $1.83\%$ ($17,278$ out of $945,767$) affiliation strings have no country name detected. The remaining $3\%$ affiliation strings either contain more than one country name or have more than 8 fields which may represent multiple affiliations.\\
 
The next step is to focus on `\textit{splitting the multiple affiliations}' into single records. The case of an affiliation string with multiple country names varies. For instance, it may represent one affiliation but include the country names with overlapped words (e.g., Mexico vs. New Mexico for \textit{string match} procedure, like \texttt{THE UNIVERSITY OF NEW MEXICO, ALBUQUERQUE NEW MEXICO} and Washington vs. Washington, D.C. for \textit{field match} procedure, like \texttt{THE GEORGE WASHINGTON UNIVERSITY, WASHINGTON,  D.C.}); or some country names may represent a city, a region or a street, (e.g., \texttt{ST. JOHN'S UNIVERSITY, JAMAICA, NEW YORK}); or the union states for some historical countries (e.g. \texttt{FACULTY OF CIVIL ENGINEERING, UNIVERSITY OF BELGRADE, BULEVAR REVOLUCIJE 73, 11000 BEOGRAD, SRBIJA, YUGOSLAVIA}). We go through this scenario first, and try to filter out affiliation strings of unique affiliation. We assume that two country names cannot appear in the neighbor fields or in the neighbor words. Thus, if we found two country names in neighboring fields, we consider the latter one as the real country name. But if two country names are in the same comma separated field, we determine the country name(s) based on their position. We assign an index to each of the words in that field according to the order of the words. If the number of words between the first indices of two country names is less than the number of the words of the longer country name, the country name with the larger length is the country name. For instance, in the above example \texttt{THE UNIVERSITY OF NEW MEXICO, ALBUQUERQUE NEW MEXICO}, we find two country names in the second field: \texttt{NEW MEXICO} and \texttt{MEXICO} with the word indices 2 and 3 respectively. The number of words between two indices is 1, which is smaller than the length of \texttt{NEW MEXICO}, so we determine \texttt{NEW MEXICO} is the country name for this affiliation.\\

After performing the multiple name checking described above, we consider the remaining affiliation strings consisting of multiple affiliations. We observe that the affiliation strings in this scenario usually contain elements implying multiplicity, like \texttt{AND} and semicolons. For example:
\begin{itemize}
\item[] \texttt{THE RICE INSTITUTE, HOUSTON, TEXAS AND THE COLLEGE OF THE PACIFIC, STOCKTON, CALIFORNIA}
\item[] \texttt{INSTITUTE FOR ADVANCED STUDY, PRINCETON, NEW JERSEY 08540 AND PHYSICS DEPARTMENT, CALIFORNIA INSTITUTE OF TECHNOLOGY, PASADENA, CALIFORNIA}
\item[] \texttt{ISTITUTO DI FISICA DELL{'}UNIVERSITA, ROMA, ITALY; AND ISTITUTO NAZIONALE DI FISICA       NUCLEARE, SEZIONE DI ROMA, ITALY}
\end{itemize}
If there are semicolons in the affiliation strings, we split the affiliation strings by the position of the semicolon. However, if there is no semicolon, while there is an \texttt{AND}, we have to exclude the case like `\texttt{DEPARTMENT OF PHYSICS AND ASTRONOMY}'. To do so, we observe that if an \texttt{AND} joins two affiliations, the country name usually should appear closely before the \texttt{AND}, so we split the string into two part by an \texttt{AND} if the last word position of the country name before \texttt{AND} is at most one word far from the \texttt{AND} (We allow one word between the country name and \texttt{AND} because of possible non-numeric postal codes.), and the \texttt{AND} does not join any two of the descriptive words of research subjects, which usually appear in the information of institute and sub-institute. We built a list of descriptive words by calculating the frequency of the word appearance in the first field of all affiliation strings. The top 20 frequently appeared descriptive words are listed in \autoref{tb:appendix:topDescriptivewords}.\\
\begin{table}[ht]
\caption{The top 20 descriptive words of research subjects.}
\centering
\scalebox{0.8}{
\begin{tabular}{ll|ll}
\hline
word & frequency & word & frequency\\
\hline
PHYSICS & 314266 & RESEARCH   &     55692\\
SCIENCE & 37345  & THEORETICAL &    32976\\
ASTRONOMY    &   32247 & ENGINEERING   &  28179\\
MATERIALS    &   27572 & PHYSIK  & 24083\\
CHEMISTRY    &   23821 &  FISICA & 23649\\
F\'{I}SICA  & 22711 & PHYSIQUE    &    21928\\
NUCLEAR & 21860 & TECHNOLOGY  &    18769\\
SCIENCES   &     16999 & APPLIED & 16184\\
THEORETISCHE  &  12994 & MATHEMATICS &    10978\\
SOLID  & 10351 & PHYSICAL    &    9194\\
\hline
\end{tabular}
}
\label{tb:appendix:topDescriptivewords}
\end{table}

For the affiliation strings with more than 7 fields, e.g.,
\begin{itemize}
\item[] \texttt{CENTER FOR THEORETICAL PHYSICS, DEPARTMENT OF PHYSICS AND ASTRONOMY, UNIVERSITY OF TEXAS AT AUSTIN, TEXAS 79712; CENTER FOR ADVANCED STUDIES, DEPARTMENT OF PHYSICS AND ASTRONOMY, UNIVERSITY OF NEW MEXICO, ALBUQUERQUE, NEW MEXICO 97131; AND MAX-PLANCK-INSTITUT F\"{U}R QUANTENOPTIK, D-8046 GARCHING BEI MUNCHEN, WEST GERMANY}
\end{itemize}
we first split it by semicolons but not by \texttt{AND}. The split substrings will be processed step by step from \textit{field match} to \textit{string match} and possibly \textit{splitting multiple affiliations}, in the same way as an affiliation string with no more than 7 fields is processed. \\

It is worth to note that even after splitting process, some of the affiliation strings still contain more than one country name, like
\begin{itemize}
\item[] \texttt{LOS ALAMOS NATIONAL LABORATORY, UNIVERSITY OF CALIFORNIA, LOS ALAMOS, NEW MEXICO}
\end{itemize}
for which the above steps give both California and New Mexico as its country names, or
\begin{itemize}
\item[] \texttt{INSTITUTE FOR QUANTUM COMPUTING, UNIVERSITY OF WATERLOO, N2L 3G1, WATERLOO, ON, CANADA, ST. JEROME{'}S UNIVERSITY, N2L 3G3, WATERLOO, ON, CANADA, AND PERIMETER INSTITUTE FOR THEORETICAL PHYSICS, N2L 2Y5, WATERLOO, ON, CANADA}
\end{itemize}
of which the first substring after splitting by \texttt{AND} (\texttt{INSTITUTE FOR QUANTUM COMPUTING, UNIVERSITY OF WATERLOO, N2L 3G1, WATERLOO, ON, CANADA, ST. JEROME{'}S UNIVERSITY, N2L 3G3, WATERLOO, ON, CANADA}) still contains another affiliation and there is no more semicolon and \texttt{AND} to indicate the position to split. \autoref{fig:appendix:fieldDistribution} shows that on average affiliation strings representing a single affiliation consist of four fields, therefore we split the affiliation (sub)strings of multiple country names but without any semicolon and \texttt{AND} at the position of the country names if the number of fields between two country names is not smaller than 4. Thus the final country names for the affiliation strings of the above two examples are `New Mexico' and three `Canada's respectively. \\

To double check the results obtained from the above procedures, we use Google geocoders from geopy toolbox \cite{geopy} to get the country names searched by Google map, and call this step \textit{Google geocoders checking}. Unfortunately, Google geocoders usually cannot code the affiliation strings with department information or even institution information. To avoid these exceptions, for the affiliation string with more than three fields, we send the last three fields as an address string to geocoders, and for others we input the whole string to geocoders. Google geocoders return a comma separated address string for each input. If the returned string is not empty, we match the country names, 2-letter or 3-letter abbreviations in our country name dataset with the returned result. Once the matched result represent the same country as we extracted, we say the country name we parsed for this affiliation string is validated. It should be noted that we do not use Google geocoders (or other geocoders like Yahoo! or Bing) directly to search country names because to our best knowledge there is no evidence to guarantee the accuracy of the results from these APIs.Thus we perform this step of checking to get better accuracy.\\

\begin{figure}[ht]
\centering
\includegraphics[width=\textwidth]{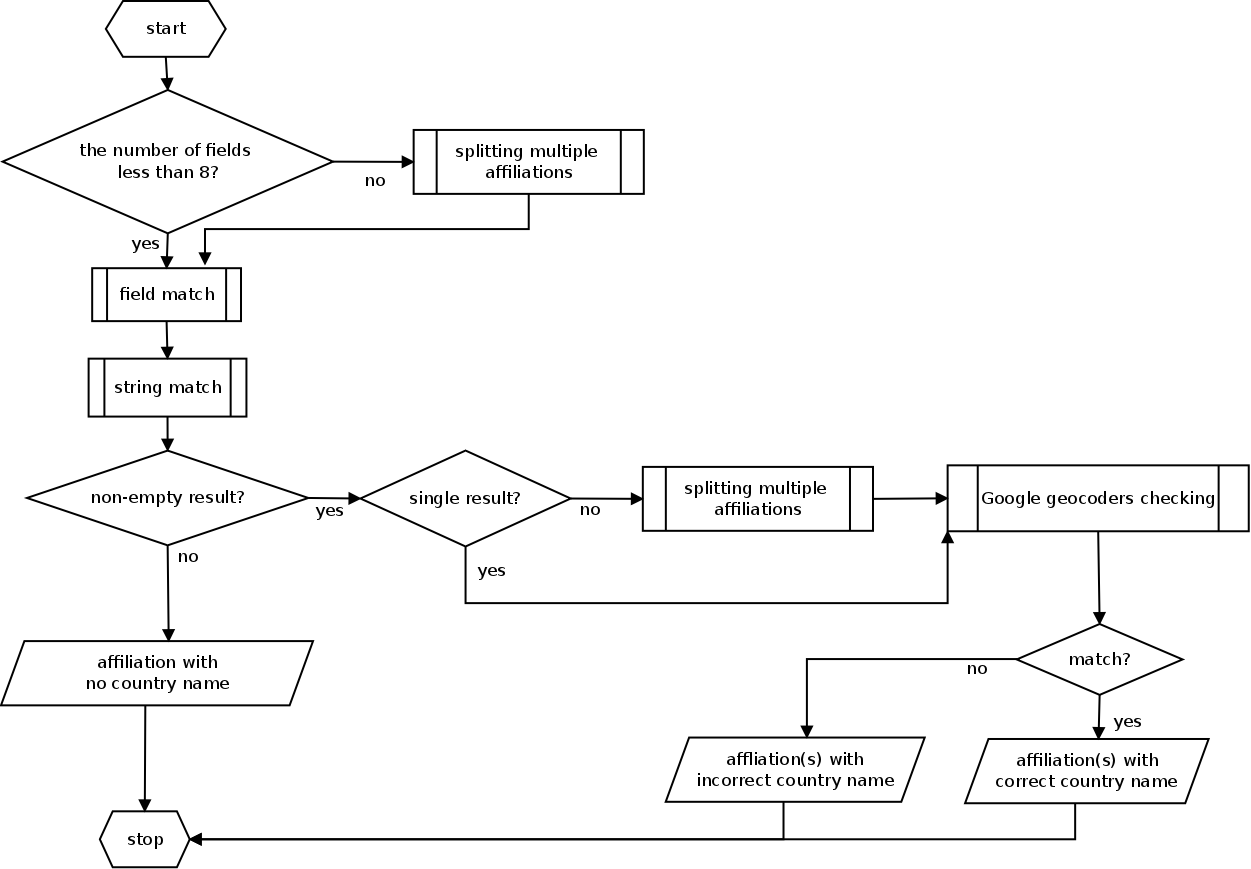}
\caption{The flow chart of the procedure to extract country name(s) from affiliation strings.}
\label{fig:appendix:dataparsingFlowchart}
\end{figure}

\begin{figure}
\centering
\includegraphics[width=0.55\textwidth]{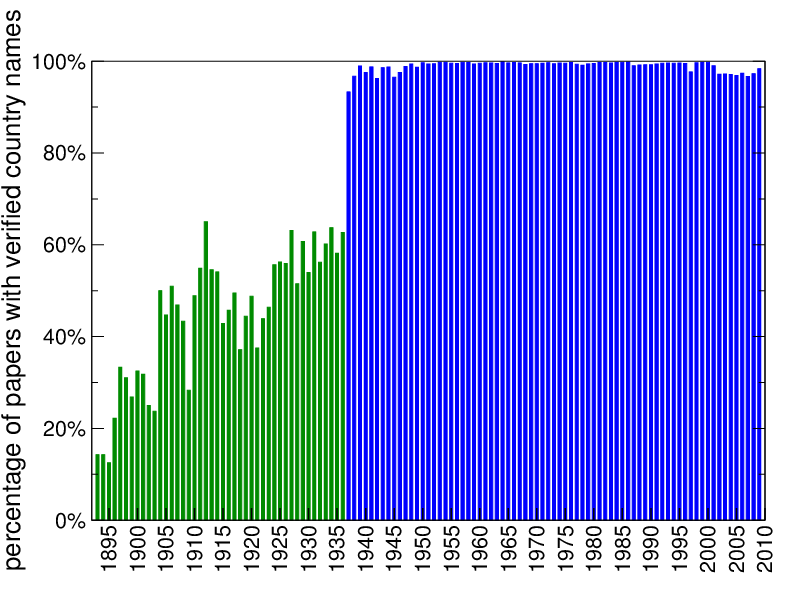}
\caption{The percentage of papers (DOIs) with validated country names per year. The plot shows that after 1940 we obtain more than $95\%$ of papers with verified country names (blue bars).}
\label{fig:appendix:countryAnalysis:overYear}
\end{figure}

\autoref{fig:appendix:dataparsingFlowchart} summarizes the above steps to extract country names from affiliation strings in a flow chart. As the result, the $3\%$ of affiliation strings with multiple country names and more than 7 fields are finally split into $46,353$ new records. In the end, we obtain $963,206$ records of single affiliation, of which $97.68\%$ ($940,896$) have a country name validated with Google geocoders. \autoref{fig:appendix:countryAnalysis:overYear} indicates that after 1940, we parsed validated country names for more than $95\%$ of papers in each year. We use these affiliation strings with validated country names to build citation networks at the country level after 1940, and as the inputs to extract city names.


\subsection{Parsing city names}
We use the database of GeoNames to parse the name of cities in the affiliation strings with identified country names. GeoNames database includes geographical data such as names of villages, cities, and other types of places in various languages, elevation, population and others from various sources \cite{GeoName}. The variations of languages for geographic names allow us to identify city names written in languages other than English. Each record of places in the database also includes its country name and possibly the first level of administrative division (e.g., the states in the United States). We first filter records that represent cities (by the feature codes attribute in GeoNames data), and arrange cities by the names of countries and US states. For countries like the Soviet Union and Yugoslavia, we combine the cities of their former union countries; and for East Germany we simply use the cities in Germany.\\

The final results from the above section is a set of affiliation strings, each of which owns a unique country name, so we argue, that to our best effort, each affiliation string now only represents an institution and has one city name if any. Since each affiliation string now has a validated country name, we only use the city list of that country to avoid the same city name in different countries. \\

After cleaning the data, the first step to parse city names is `\textit{field match}', as we performed to find country names. For each field, we delete words with numbers and try to match it with city names in filtered city dataset for that country. If there are matched city names, we list both the name and coordinates as outputs, otherwise we perform `\textit{string match}' on the affiliation strings trying to match city names word by word.\\

As we did to validate country names, we use Google geocoders from geopy toolbox to check the correctness of the city names we extract from affiliation strings. The procedure is similar to that for the country names: the affiliation strings excluding the department level information are given as input to Google geocoders, and the non-empty Google searched results are saved for the next step of validation.The coordinates and city names given by Google geocoders for an affiliation string are based on the name of the institutions, and may be different from the name extracted and the coordinates of the city given in GeoName database. To determine if the extracted city name is correct, we simply calculate the geographic distance between the coordinates given by GeoNames database and the ones given by Google geocoders, and if the distance is less than 50km, we say the extracted result is matched with Google searched result. For the affiliation strings with multiple city names, we choose the one which has the shortest Vincenty's distance from the Google geocoded result.\\

\begin{figure}[h]
 \begin{center}
  \subfloat[The percentage of papers with validated city names per year.]{\label{fig:appendix:cityAnalysis:overYear}\includegraphics[width=0.5\textwidth]{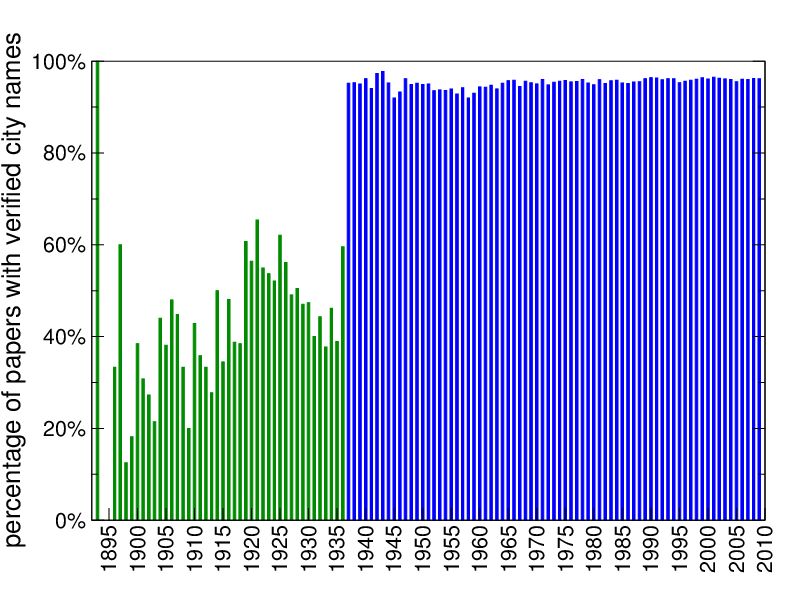}}
  \subfloat[The percentage of papers with validated city names per country.]{\label{fig:appendix:cityAnalysis:overCountry}\includegraphics[width=0.5\textwidth]{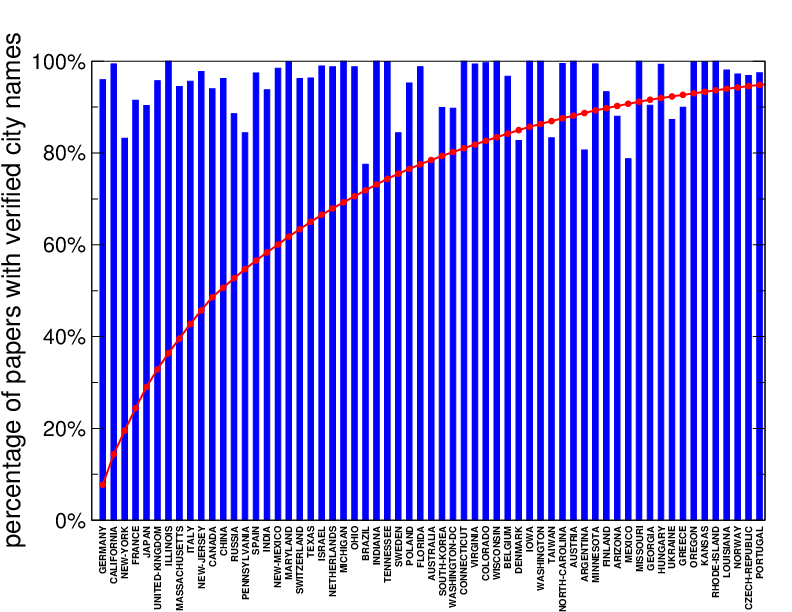}}
 \end{center}
\caption{The percentage of papers (DOIs) with validated city names per year \newsubref{fig:appendix:cityAnalysis:overYear}, and the percentage of papers (DOIs) with validated city names per country \newsubref{fig:appendix:cityAnalysis:overCountry}. \newsubref{fig:appendix:cityAnalysis:overYear} clearly shows that after 1940 we obtain more than $90\%$ of papers with verified city names for each year (blue bars). In \newsubref{fig:appendix:cityAnalysis:overCountry}, the x-axis is top 60 countries ranked by the total number of papers after 1940 in each country. The red dot curve is the cumulative distribution function of the number of papers over countries after 1940. For the major contributing countries in terms of paper production, we have obtained more than $80\%$ of papers with validated city names.}
\label{fig:appendix:cityAnalysis}
\end{figure}

In total, we have $92.6\%$ ($871,345$ out of $940,896$) affiliation strings with validated city names. \autoref{fig:appendix:cityAnalysis:overYear} shows the the percentage of papers (DOIs) with validated city names per year, from which one can observe that we obtain validated city names for more than $90\%$ of papers after 1940, and for this reason we use data after that year to perform analysis at the city level in this paper. \autoref{fig:appendix:cityAnalysis:overCountry} displays the percentage of papers with validated city names to the total number of papers for each country after 1940. The abscissa is 60 country names ordered by the total number of papers for each country after 1940. These top 60 countries contribute $95\%$ of the papers published in Physical Review journals after 1940, as shown by the cumulative distribution of the total number of papers for all countries (the red dot curve). From \autoref{fig:appendix:cityAnalysis:overCountry} we claim that for the most of major countries contributing to publications in Physical Review journals we have unbiased results of parsing city names.\\

So far we have obtained geographic coordinates and city names for the affiliation strings from Google geocoders and GeoName database. However, different city names may represent the same city, geographically close cities or different administrative levels. For instance,
\begin{itemize}
\item[] \texttt{DEPARTMENT OF PHYSICS, BOSTON COLLEGE, BOSTON, MASSACHUSETTS 02467, USA}
\item[] \texttt{DEPARTMENT OF PHYSICS, BOSTON COLLEGE, CHESTNUT HILL, MASSACHUSETTS}
\end{itemize}
Because Chestnut Hill is not a city in Massachusetts in GeoNames database, the city name extracted from these two affiliation strings for Boston College is Boston, while Google geocoders gives the city name of Newton. In this case, one cannot automatically determine which city this affiliation should be in. One possible way to solve such the problem is to project the coordinates into polygons of `cities' in shapefiles for geographic information systems software. However, the existent shapefiles have different granularities for different countries. It may be unfair to compare the scientific products in different level of administrative units over different countries.\\

Therefore, we cluster cities according to their geographic coordinates into `urban areas' or `academic cities' in each country. For each country, we perform hierarchical/agglomerative clustering with the geographic distance matrix, of which the distances are calculated with Vincenty's formula. With the dendrogram produced from the clustering process, we cut off the branches from the maximum height value to lower ones until the distance between any point in a cluster and the centroid of the cluster is less than 25km (the maximum distance within the cluster is 50km) for all clusters. We call such clusters `academic cities'. The final coordinates of an academic city is the centroid of all coordinates inside that cluster, and the academic city is named with the city name which has the most papers in that cluster. We notice that due to the differences between geographic areas in different countries, some cities are merged into one academic city and some other cities are split into two. For instance, Boston, Cambridge, Newton in Massachusetts are now clustered into one urban area with the name Boston; and Dubna in Moscow Oblast now becomes a separate academic city. Finally, we have a list of academic cities for each paper (DOI), and all the analysis we made at the city level in this paper refer to the unban areas or academic cities.

\section{Building the citation networks}
A citation network consists of a set of nodes (cities) and directed links representing citations that one paper written in one city is cited by a paper written in another one according to the references of the latter. For example, if a paper is written in node $i$ cites one paper written in node $j$ there is an edge from $i$ to $j$, i.e., $j$ receives a citation from $i$ and $i$ sends a citation to $j$. As shown in Figure (1) in the main text, a directed link from Ann Arbor to Rome and another link to Madrid are built since paper \textit{A}, which is from Ann Arbor, Michigan, cites the paper \textit{B} from Rome, Italy and Madrid, Spain. Because the paper \textit{A} was also contributed by authors from another two cities: Los Alamos in New Mexico and New York City in New York, from each of these two cities, there is also a link to Rome and another to Madrid. \\

The weight of a link is defined as following. In a given time window, the total number of citations for the papers written in $j$ received from papers written in $a$, is the weight of the link $(i \rightarrow j)$, and the total number of citations for those paper written in $j$ sent to the papers written in $k$ is the weight of the link $(j \rightarrow k)$. For instance, in time window $t$, there is one paper written in node $j$, which cited two papers written in node $k$ and was cited by three papers written in node $i$, then there are $w_{i,j}=3, w_{j,k}=2$, and we add up such weight for all papers written in that node $j$ and obtain the weights for links. For the paper written in multiple cities, say $j_1,j_2$, the weight will be counted equally, i.e., $w_{i,j_1} = w_{i,j_2}, w_{j_1,k} = w_{j_2,k}$. The time window we use in this paper is 1 year.

\section{Basic properties of data and citation networks}
\begin{figure}[h]
\centering
\begin{minipage}{0.5\textwidth}
\centering
\captionsetup{width=.95\textwidth} 
\includegraphics[width=.95\textwidth]{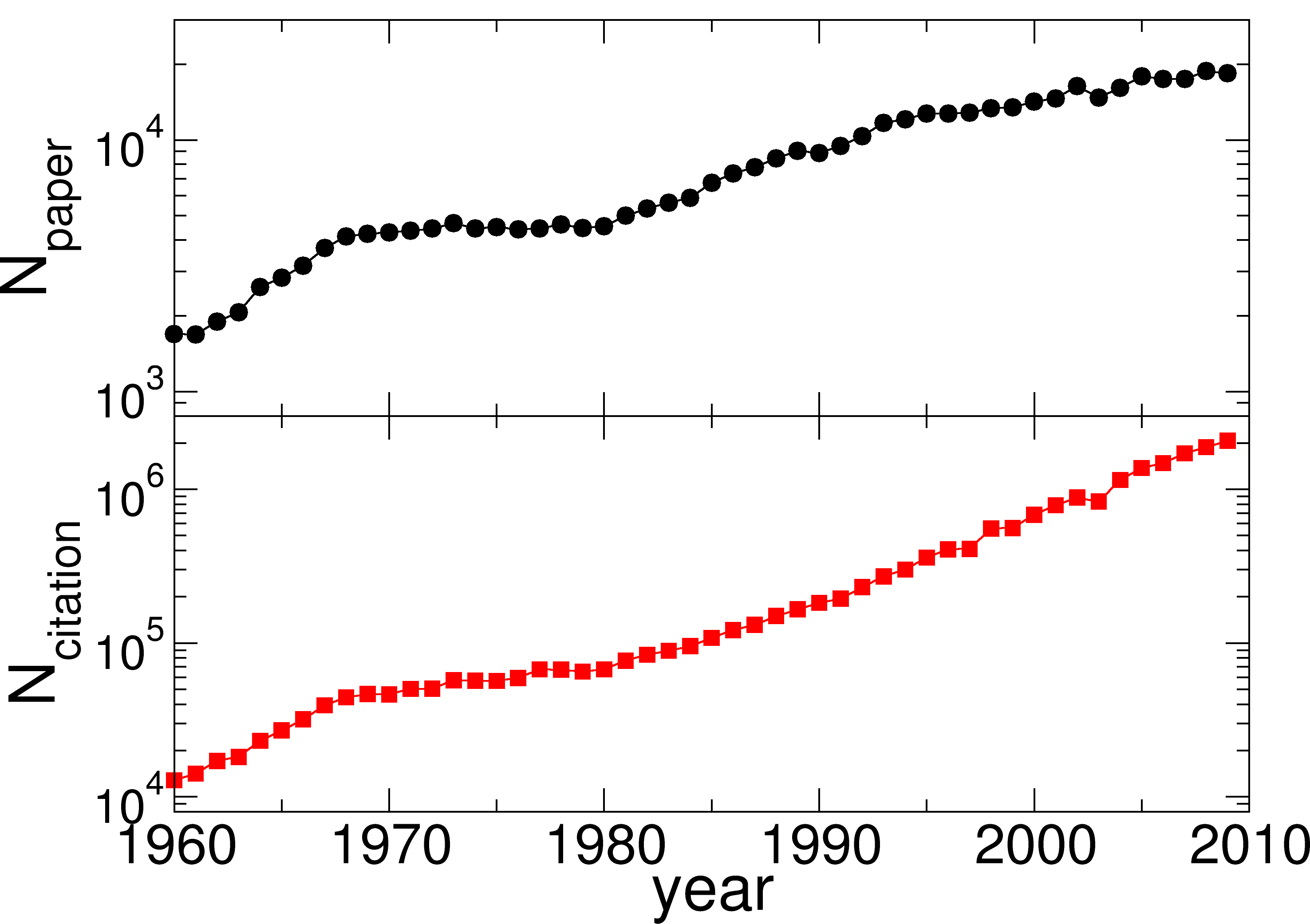}
\caption{The number of papers (top) and the number of citations (bottom) as the function of time (1960-2009).}
\label{fig:appendix:NpaperNcitation}
\end{minipage}%
\begin{minipage}{0.5\textwidth}
\centering
\captionsetup{width=.95\textwidth}
\includegraphics[width=.95\textwidth]{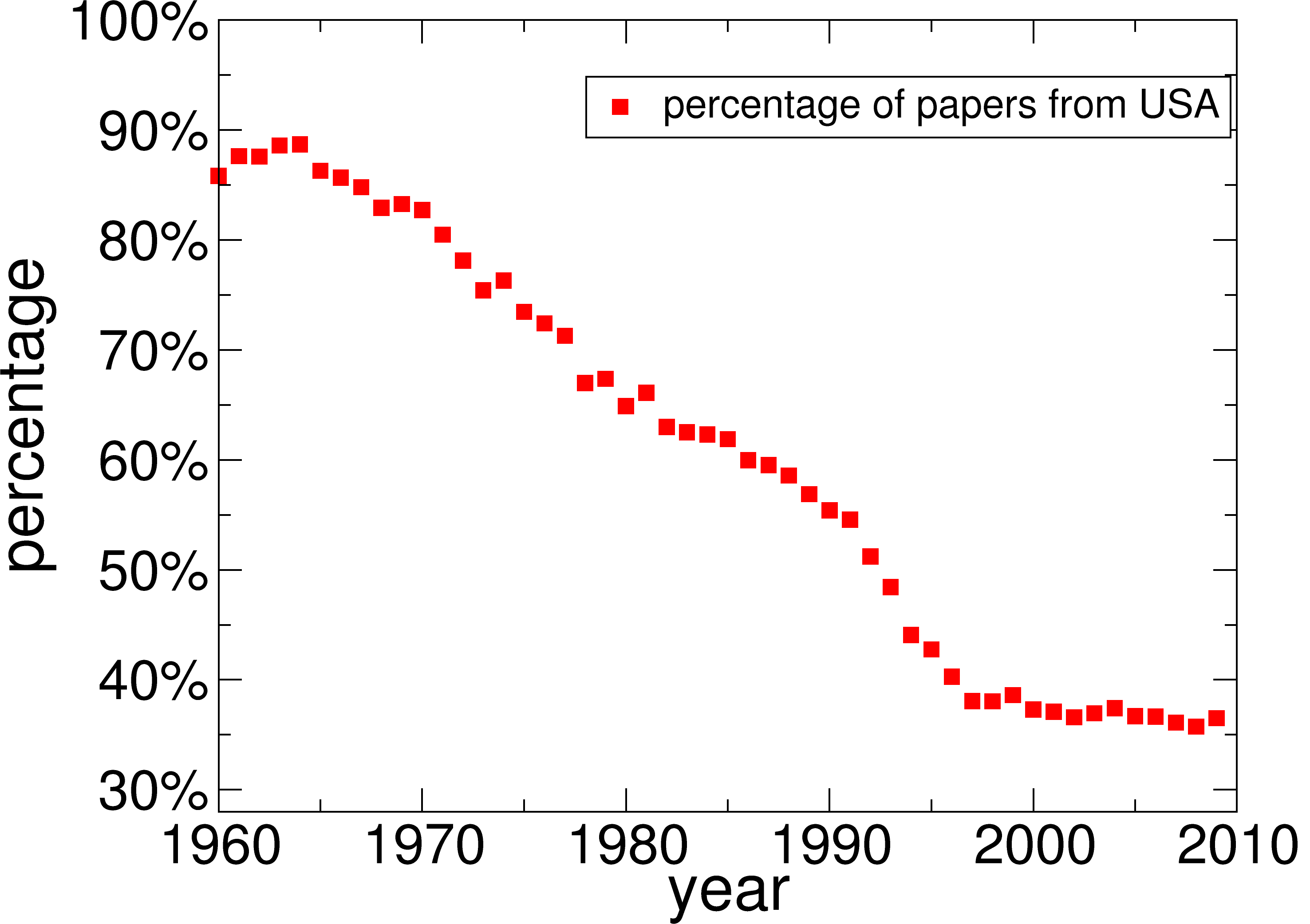}
\caption{The percentage of papers contributed by authors from USA as the function of time (1960-2009).}
\label{fig:appendix:USratio}
\end{minipage}
\end{figure}

\begin{figure}[h]
\centering
\begin{minipage}{0.5\textwidth}
\centering
\captionsetup{width=.95\textwidth} 
\includegraphics[width=.95\textwidth]{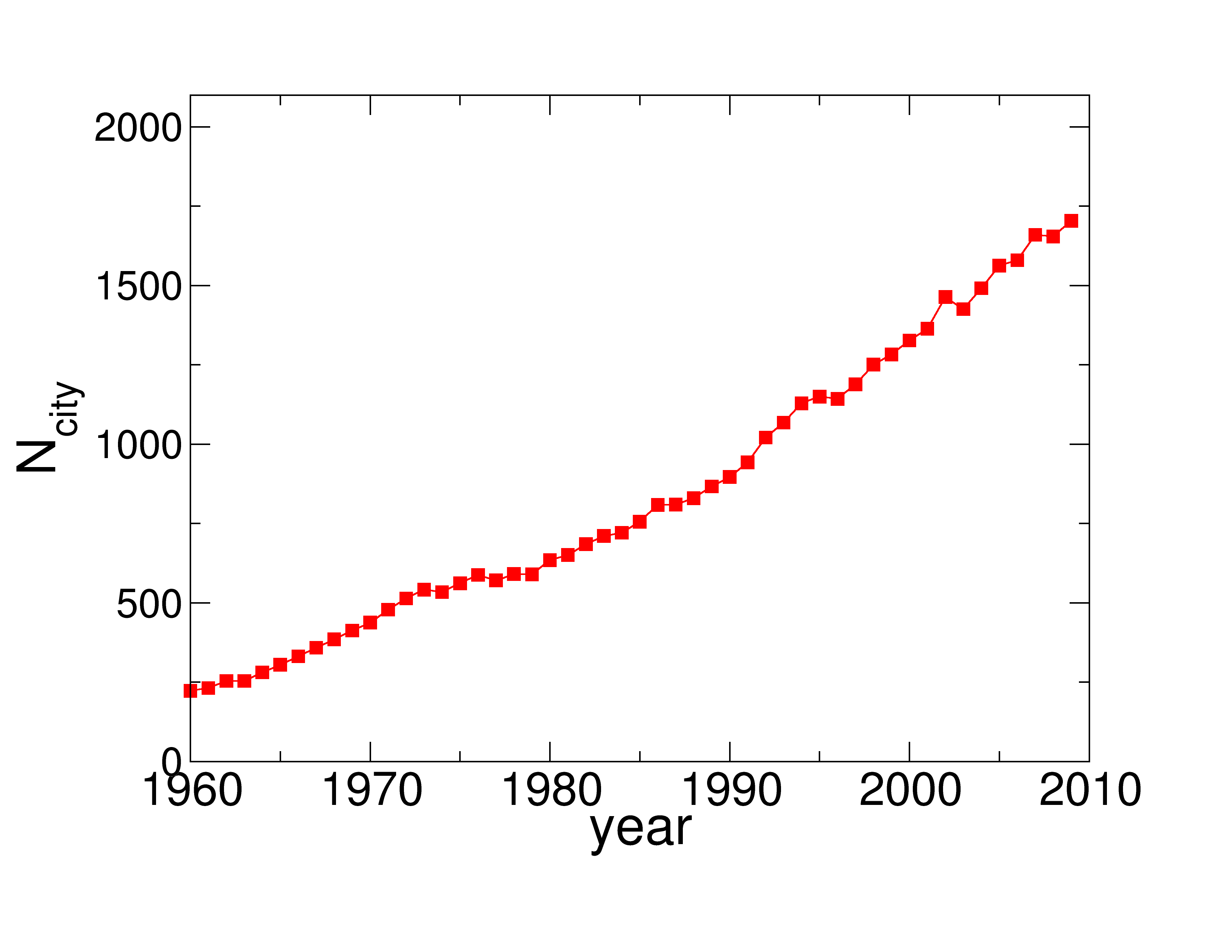}
\caption{The number of nodes (cities) for city-to-city citation networks as the function of time (1960-2009).}
\label{fig:appendix:Ncity}
\end{minipage}%
\begin{minipage}{0.5\textwidth}
\centering
\captionsetup{width=.95\textwidth}
\includegraphics[width=.95\textwidth]{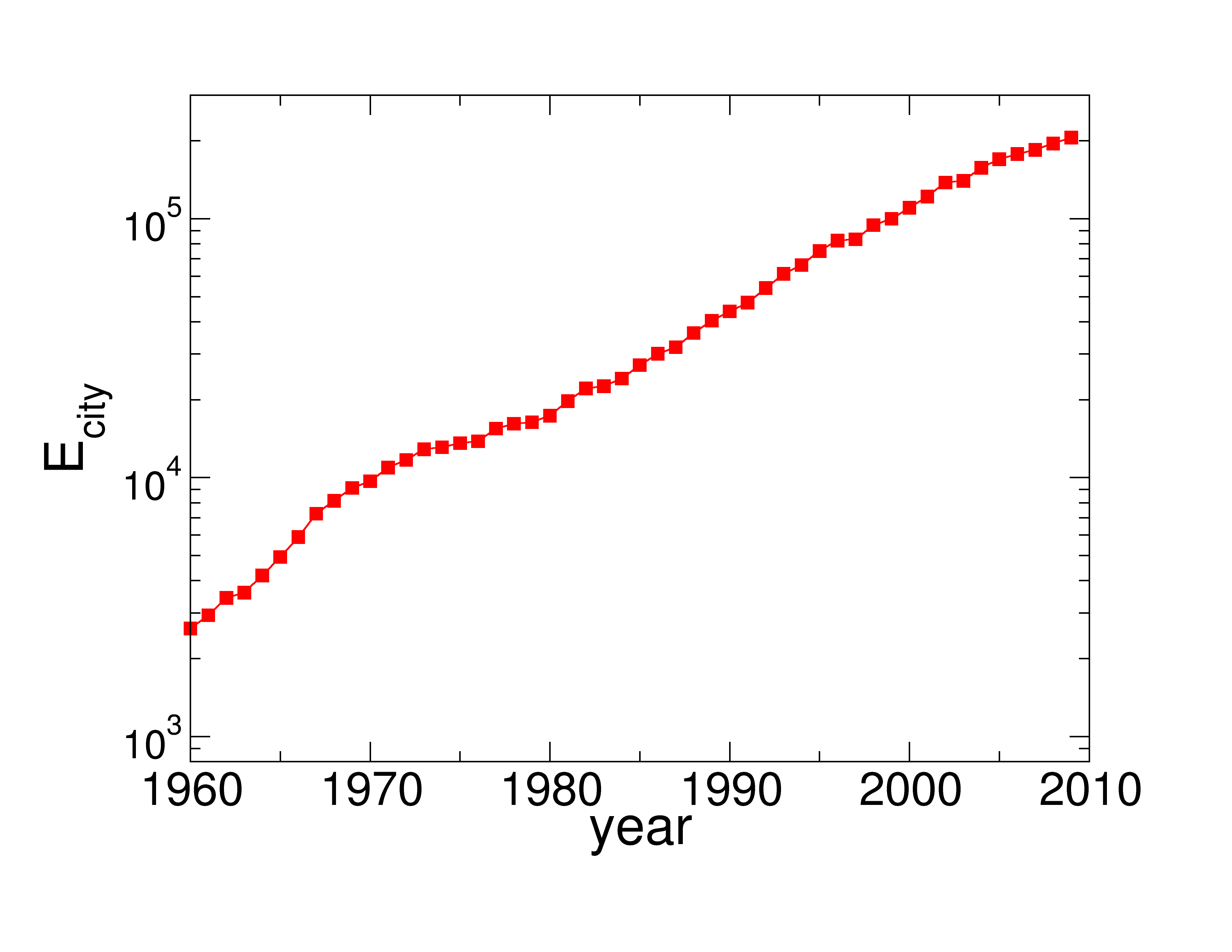}
\caption{The number of links for city-to-city citation networks as the function of time (1960-2009).}
\label{fig:appendix:Ecity}
\end{minipage}
\end{figure}

We observe a significant growth of the published articles and the citations in recent $50$ years, as shown in \autoref{fig:appendix:NpaperNcitation}. Meanwhile, the percentage of papers contributed by authors in the United States has decreased from nearly $90\%$ in early 1960's to current $36\%$ (\autoref{fig:appendix:USratio}).  Correspondingly, the number of cities contributing to publications in APS journals, as well as their internal interactions, has increased dramatically, as illustrated in \autoref{fig:appendix:Ncity} and \autoref{fig:appendix:Ecity}. \\

\begin{figure}[h]
\begin{center}
\captionsetup{width=.4\textwidth}
\subfloat[The cumulative distribution function of the degrees for citation networks at the city level.]{\label{fig:appendix:distribution:citydegree}\includegraphics[width=0.45\textwidth]{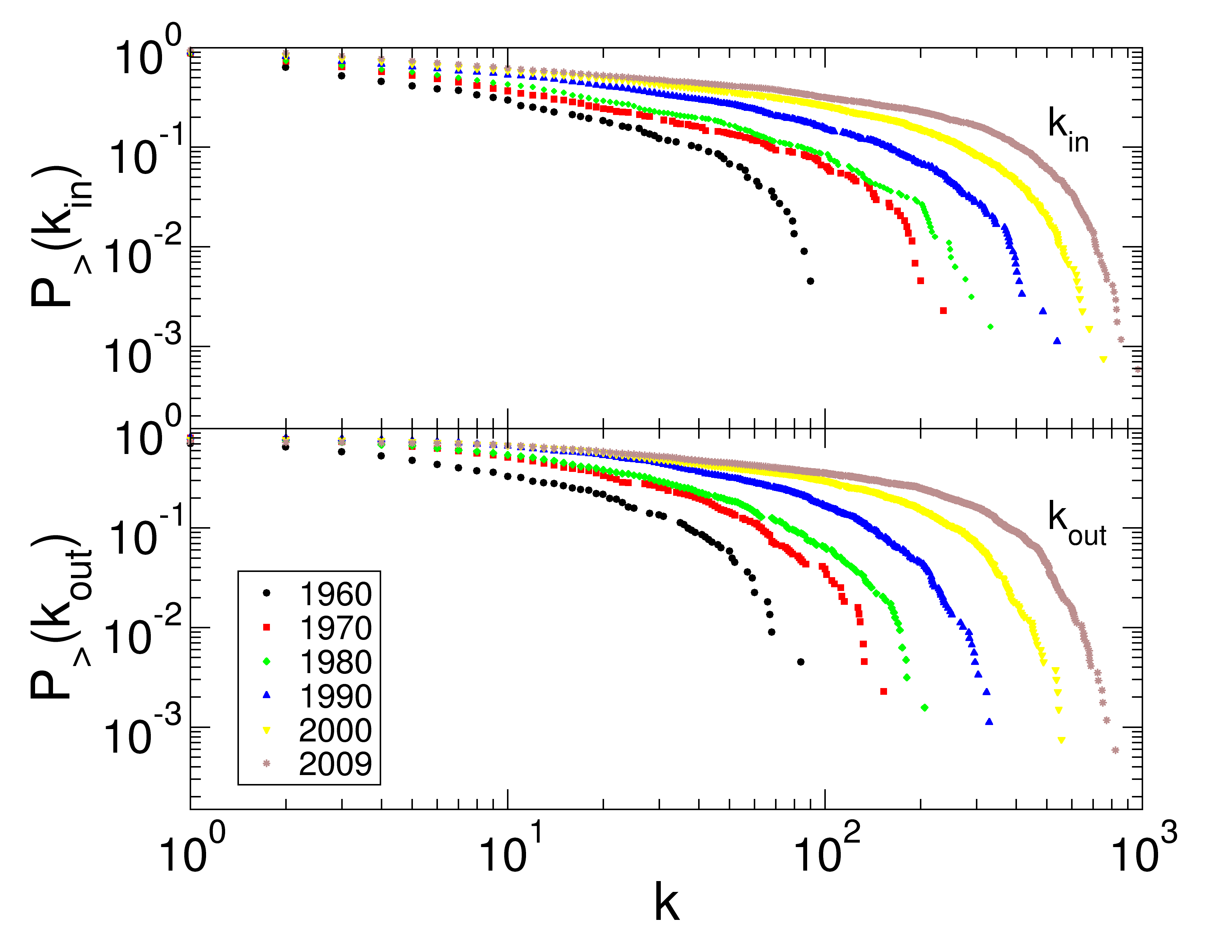}}
\subfloat[The cumulative distribution function of the strength for citation networks at the city level.]{\label{fig:appendix:distribution:citystrength}\includegraphics[width=0.45\textwidth]{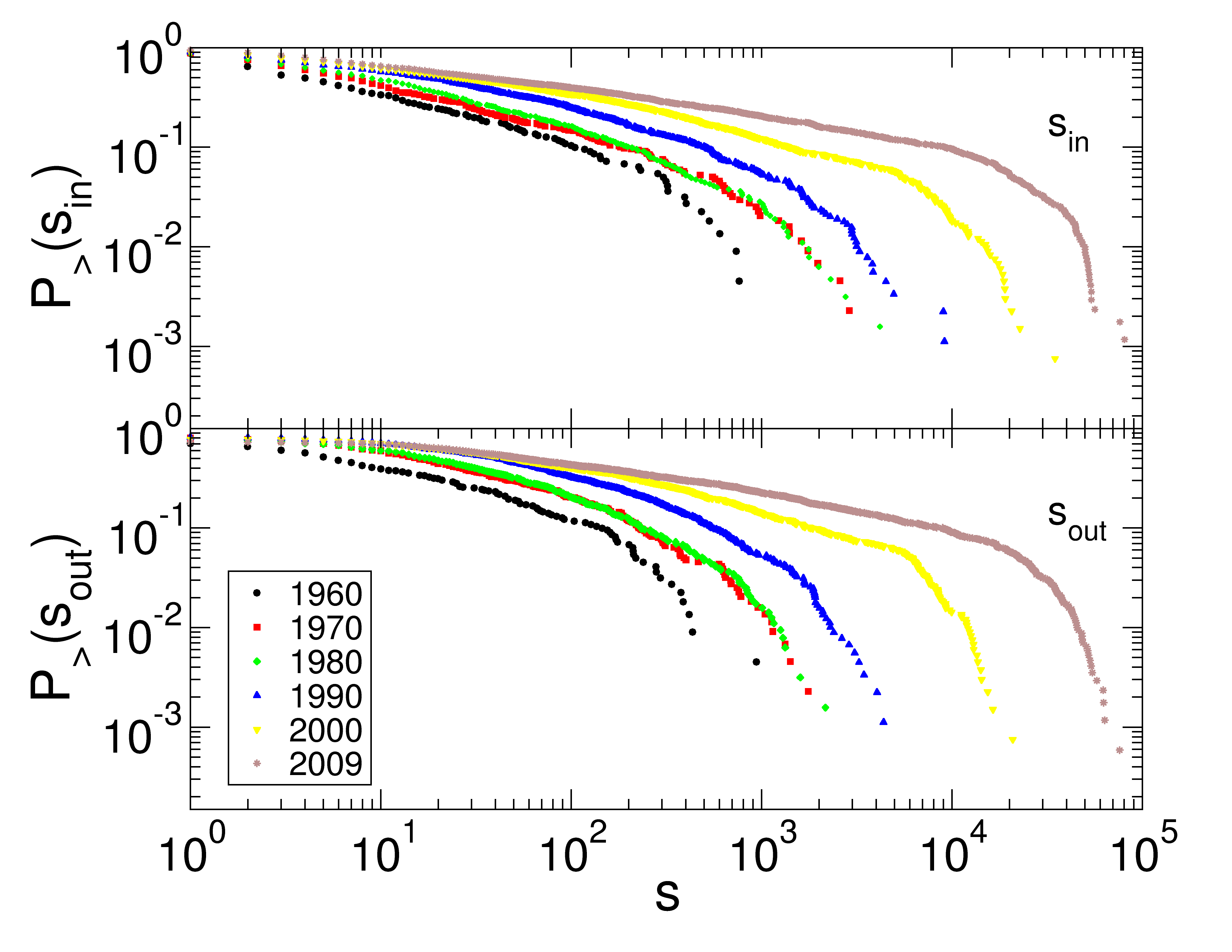}}
\end{center}
 \caption{The cumulative distribution function of degree and strength for city-to-city citation networks in year 1960, 1970, 1980, 1990, 2000 and 2009.}
 \label{fig:appendix:distribution}
\end{figure}

In \autoref{tb:appendix:cityNetSummary} we report basic statistic properties for the city-to-city citation networks in selected years. \autoref{fig:appendix:distribution:citydegree} reports the cumulative distribution functions for in- and out-degree of the city-to-city citation networks in different years. The distributions are with behaviors close to power-law with the exponential cutoff. As the year increases, the range of values of $k_{\mathrm{in}}$ and $k_{\mathrm{out}}$ extends. We define the in/out-strength of node $i$ as the total number of citations it sends/receives at that year. \autoref{fig:appendix:distribution:citystrength} displays the cumulative distribution function for in- and out-strength of the city-to-city citation networks in different years. The pattern of strength distributions is quite similar to the degree distributions. 

\begin{table}[ht]
\centering
\caption{Summary of basic statistic features for city-to-city citation networks in different years.}
\label{tb:appendix:cityNetSummary}
\scalebox{0.55}{
\begin{tabular}{c|c|c|cccc|cccc|cccc|cccc|cccc}
\hline
  \multirow{2}{*}{year}  & \multirow{2}{*}{$V$} & \multirow{2}{*}{$E$} & \multicolumn{4}{c|}{$k_{\mathrm{in}}$} & \multicolumn{4}{c|}{$k_{\mathrm{out}}$} & \multicolumn{4}{c|}{$S_{\mathrm{in}}$} & \multicolumn{4}{c|}{$S_{\mathrm{out}}$} & \multicolumn{4}{c}{$w_{ij}$} \\
  \cline{4-23}
    &   &  & mean & std. & min & max & mean & std. & min & max & mean & std. & min & max & mean & std. & min & max & mean & std. & min & max\\
\hline
1960 & 222 & 2517 & 11.34 & 18.13 & 0 & 90 & 11.34 & 15.20 & 0 & 84 & 41.24 & 111.16 & 0 & 765 & 41.24 & 95.99 & 0 & 940 & 3.64 & 11.57 & 1 & 336\\
1970 & 438 & 9461 & 21.60 & 38.97 & 0 & 236 & 21.60 & 26.72 & 0 & 153 & 87.53 & 288.39 & 0 & 2893 & 87.53 & 198.54 & 0 & 1758 & 4.05 & 13.98 & 1 & 564\\
1980 & 635 & 17028 & 26.82 & 47.96 & 0 & 332 & 26.82 & 34.84 & 0 & 206 & 94.08 & 311.71 & 0 & 4182 & 94.08 & 213.94 & 0 & 2164 & 3.51 & 11.02 & 1 & 557\\
1990 & 897 & 43324 & 48.30 & 80.31 & 0 & 539 & 48.30 & 58.37 & 0 & 329 & 207.59 & 671.95 & 0 & 9125 & 207.59 & 459.34 & 0 & 4372 & 4.30 & 13.00 & 1 & 830\\
2000 & 1327 & 109438 & 82.47 & 126.79 & 0 & 754 & 82.47 & 102.83 & 0 & 556 & 801.76 & 2640.94 & 0 & 34768 & 801.76 & 2167.73 & 0 & 20862 & 9.72 & 29.71 & 1 & 1568\\
2009 & 1704 & 204747 & 120.16 & 178.22 & 0 & 968 & 120.16 & 151.16 & 0 & 822 & 3033.86 & 9230.21 & 0 & 104149 & 3033.86 & 8651.34 & 0 & 76044 & 25.25 & 75.12 & 1 & 3004\\
\hline
\end{tabular}}
\end{table}

\clearpage
\section{Top producers/consumers and results from knowledge diffusion proxy}
\begin{figure}[h]
\begin{center}
\captionsetup{width=.42\textwidth}
\includegraphics[width=0.5\textwidth]{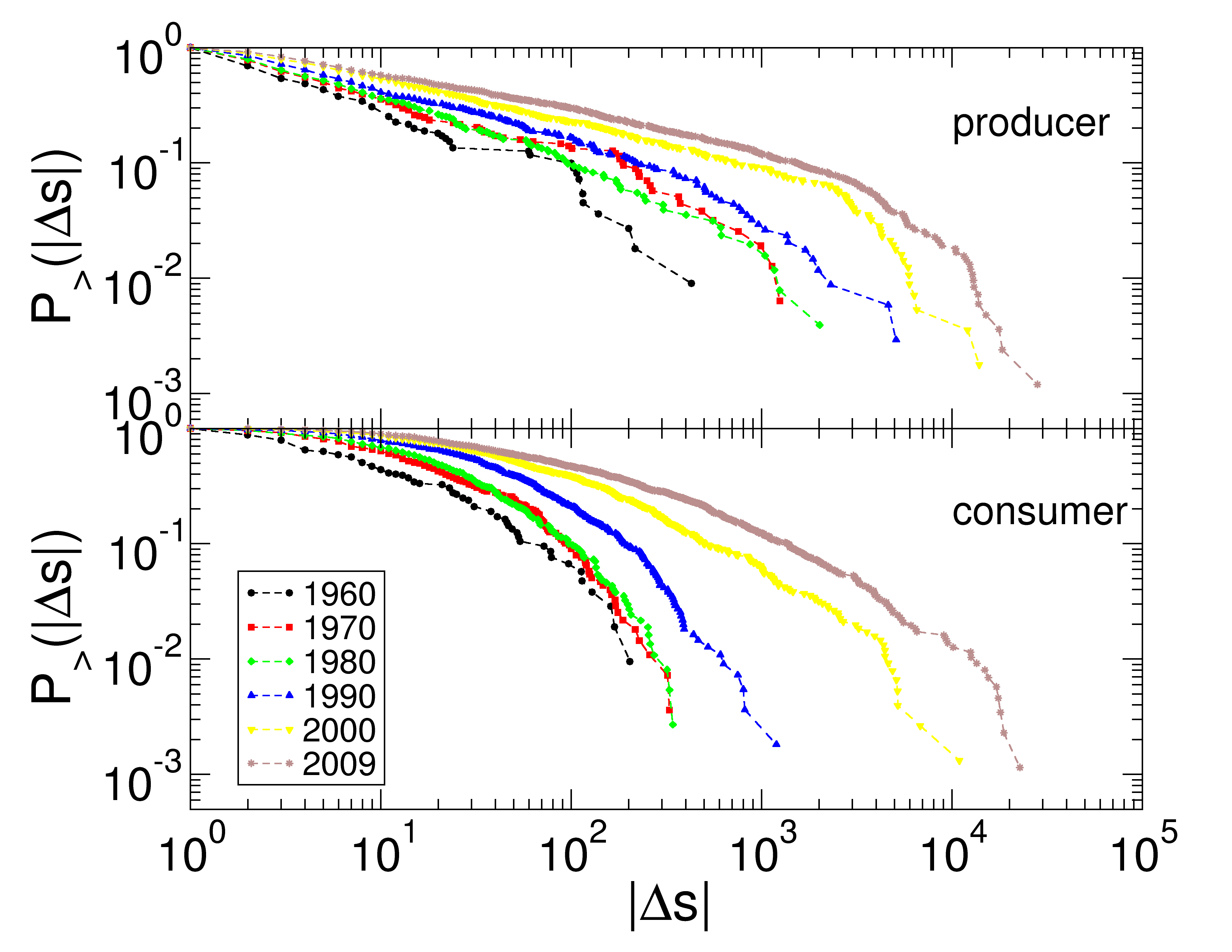}
\end{center}
 \caption{The cumulative distribution function of the citation unbalance for producers and consumers at the city level in year 1960, 1970, 1980, 1990, 2000 and 2009.}
 \label{fig:appendix:deltastrength}
\end{figure}

In \autoref{fig:appendix:deltastrength} we show the cumulative distribution of the absolute citation unbalance $|\varDelta s|$ for producers and consumers at the city level. Similar to the cumulative distributions of strength, the distributions are characterized with heavy tails, and the distributions have become broader as the time increases. \\

We list top 20 producers and consumers at the city level from 1985 to 2009 (\autoref{tb:appendix:cityProducerConsumer1985}), from 1960 to 1980 (\autoref{tb:appendix:cityProducerConsumer1960}). It is worth noting that the definition of unbalance $\varDelta s$ is from the difference between the number of citations sent and received, which cannot distinguish between cities with a large amount of production and consumption and those with less production and consumption. \\

\begin{table}
\centering
\caption{Top 20 producers and consumers at the city level (1985-2009)}
\label{tb:appendix:cityProducerConsumer1985}
\subfloat[Top 20 producer cities]{
\label{}
\scalebox{0.8}{
\begin{tabular*}{1.2\textwidth}{@{\extracolsep{\fill}}l*{6}{l}}
\hline
rank & 1985 & 1990 & 1995 & 2000 & 2005 & 2009\\
\hline
1 & Piscataway & Piscataway & Piscataway & Boston & Boston & Boston\\
2 & Boston & Boston & Boston & Piscataway & New York City & Berkeley\\
3 & Berkeley & Palo Alto & Yorktown Heights & Los Angeles & Los Angeles & New Haven\\
4 & Princeton & Yorktown Heights & Berkeley & Berkeley & Tallahassee & Suwon\\
5 & Yorktown Heights & Berkeley & Los Angeles & Chicago & Palo Alto & Princeton\\
6 & Ithaca & Princeton & Urbana & New York City & Berkeley & Piscataway\\
7 & New York City & Ithaca & New York City & Lemont & Piscataway & Higashihiroshima\\
8 & DC & New York City & Chicago & Urbana & Urbana & Prairie View\\
9 & Palo Alto & San Diego & Ithaca & Philadelphia & Pavia & Los Angeles\\
10 & Lemont & Philadelphia & Lemont & Princeton & West Lafayette & Lubbock\\
11 & Los Angeles & Chicago & Princeton & West Lafayette & Ithaca & Palo Alto\\
12 & Chicago & Santa Barbara & Palo Alto & Batavia & Rochester & Batavia\\
13 & San Diego & Pittsburgh & Santa Barbara & Rochester & Honolulu & New York City\\
14 & Seattle & Lemont & Philadelphia & Yorktown Heights & Batavia & Nashville\\
15 & Rehovot & Los Angeles & Minneapolis & Palo Alto & Yorktown Heights & Bristol\\
16 & New Haven & New Haven & San Diego & Dallas & Irvine & Rochester\\
17 & Urbana & Orsay & Batavia & Tsukuba & Lemont & Urbana\\
18 & Pittsburgh & Holmdel & Zurich & Waltham & Minneapolis & Daegu\\
19 & Villigen & Stony Brook & Waltham & Madison & Philadelphia & Tallahassee\\
20 & Waltham & Batavia & Madison & East Lansing & Boulder & Pittsburgh\\
\hline
\end{tabular*}
}
}

\subfloat[Top 20 consumer cities]{
\label{}
\scalebox{0.8}{
\begin{tabular*}{1.2\textwidth}{@{\extracolsep{\fill}}l*{6}{l}}
\hline
rank & 1985 & 1990 & 1995 & 2000 & 2005 & 2009\\
\hline
1 & Stuttgart & Tokyo & Moscow & Beijing & Beijing & Athens\\
2 & Toronto & Beijing & Beijing & Seoul & Barcelona & Gwangju\\
3 & Gaithersburg & Tsukuba & Seoul & Lancaster & Coventry & Bratislava\\
4 & Annandale & Tallahassee & East Lansing & Grenoble & Valencia & Vancouver\\
5 & Bloomington & Vancouver & Lubbock & Dubna & Perugia & Madrid\\
6 & Minneapolis & Grenoble & Montreal & Manhattan & Moscow & Berlin\\
7 & Warsaw & Seoul & Tallahassee & Quito & Heidelberg & Trieste\\
8 & Berlin & Kolkata & Davis & Suwon & London & Mainz\\
9 & Vancouver & Charlottesville & Dallas & Stillwater & Dubna & Waco\\
10 & Ames & Durham & Taipei & Santander & Riverside & Paris\\
11 & West Lafayette & Buffalo & Berlin & Lawrence & Amsterdam & Valencia\\
12 & Charlottesville & Warsaw & Tokyo & Krak{\'o}w & Hefei & Coventry\\
13 & Seoul & Tempe & Toyonaka & Marseille & Dresden & Moscow\\
14 & Montreal & Berlin & Delhi & Tokyo & Bellaterra & Bellaterra\\
15 & Trieste & Madrid & Trieste & Karlsruhe & Shanghai & Lanzhou\\
16 & Kyoto & Sao Paulo & St Petersburg & Daegu & Evanston & Shanghai\\
17 & Tokyo & Taipei & Dresden & Udine & Taipei & Sao Paulo\\
18 & Varanasi & Brussels & Bologna & Oxford & Glasgow & Kolkata\\
19 & Rio De Janeiro & Mainz & Munich & Moscow & Liverpool & Clermont\\
20 & Ridgefield & Davis & Cambridge & Ruston & Bari & Hefei\\
\hline
\end{tabular*}
}
}

\end{table}

\begin{table}
\centering
\caption{Top 20 producers and consumers at the city level (1960-1980)}
\label{tb:appendix:cityProducerConsumer1960}
\subfloat[Top 20 producer cities]{
\label{}
\scalebox{0.8}{
\begin{tabular*}{1.2\textwidth}{@{\extracolsep{\fill}}l*{5}{l}}
\hline
rank & 1960 & 1965 & 1970 & 1975 & 1980\\
\hline
1 & Boston & Princeton & Berkeley & Boston & Boston\\
2 & Princeton & Berkeley & Boston & Berkeley & Princeton\\
3 & Urbana & Boston & Princeton & Palo Alto & Piscataway\\
4 & Oak Ridge & Piscataway & Chicago & Princeton & Berkeley\\
5 & Piscataway & New York City & Piscataway & Piscataway & Palo Alto\\
6 & New York City & Los Angeles & Palo Alto & Ithaca & Ithaca\\
7 & Los Angeles & Los Alamos & Albany & Chicago & New York City\\
8 & Los Alamos & Albany & San Diego & Oak Ridge & Chicago\\
9 & Chicago & Ann Arbor & Madison & San Diego & San Diego\\
10 & Ithaca & Pittsburgh & New York City & New Haven & Los Angeles\\
11 & Rochester & Meyrin & Pittsburgh & Los Angeles & Stony Brook\\
12 & DC & Waltham & Waltham & Urbana & New Haven\\
13 & Madison & Urbana & Meyrin & Pittsburgh & Philadelphia\\
14 & Bloomington & Cambridge & Ithaca & Batavia & Albany\\
15 & Utrecht & Bloomington & Cambridge & Providence & Urbana\\
16 & Durham & Lemont & Los Angeles & Albany & Albuquerque\\
17 & London & Ithaca & Los Alamos & Durham & Waltham\\
18 & Saskatoon & DC & New Haven & Rochester & Batavia\\
19 & Sydney & Chicago & Livermore & Livermore & College Park\\
20 & St Louis & Zurich & London & DC & Pittsburgh\\
\hline
\end{tabular*}
}
}

\subfloat[Top 20 consumer cities]{
\label{}
\scalebox{0.8}{
\begin{tabular*}{1.2\textwidth}{@{\extracolsep{\fill}}l*{5}{l}}
\hline
rank & 1960 & 1965 & 1970 & 1975 & 1980\\
\hline
1 & Berkeley & West Lafayette & Evanston & Stony Brook & Austin\\
2 & Palo Alto & Palo Alto & West Lafayette & Grenoble & Boulder\\
3 & New Haven & Orsay & Austin & Columbus & Tokyo\\
4 & Pittsburgh & College Park & Trieste & Stuttgart & Haifa\\
5 & Waltham & Albuquerque & Columbus & Toronto & Toronto\\
6 & San Diego & Livermore & Delhi & Austin & Bhubaneswar\\
7 & Lemont & Delhi & Amherst & East Lansing & Rehovot\\
8 & Livermore & Minneapolis & Rochester & Amherst & Ottawa\\
9 & West Lafayette & Trieste & Milwaukee & Mumbai & Paris\\
10 & Poughkeepsie & Providence & Baton Rouge & Denton & Santa Barbara\\
11 & Evanston & Ames & Buffalo & Mexico City & Houston\\
12 & Tallahassee & Rochester & Seattle & Munich & Golden\\
13 & Columbus & Evanston & Salt Lake City & Paris & Stuttgart\\
14 & Canberra & San Diego & Haifa & Honolulu & Kolkata\\
15 & Yorktown Heights & Syracuse & Hoboken & Montreal & Toyonaka\\
16 & Arlington & Rehovot & Lincoln & Orsay & Kyoto\\
17 & Rome & Hoboken & Gainesville & Roskilde & Grenoble\\
18 & Meyrin & Oxford & Tucson & Madison & J{\"u}lich\\
19 & Ames & El Segundo & Bloomington & West Lafayette & Vancouver\\
20 & Irvine & Milan & East Lansing & Rehovot & Kingston\\
\hline
\end{tabular*}
}
}
\end{table}
\clearpage
\section{Top ranked cities from scientific production ranking algorithm}
\begin{figure}[h]
\centering
\includegraphics[width=0.5\textwidth]{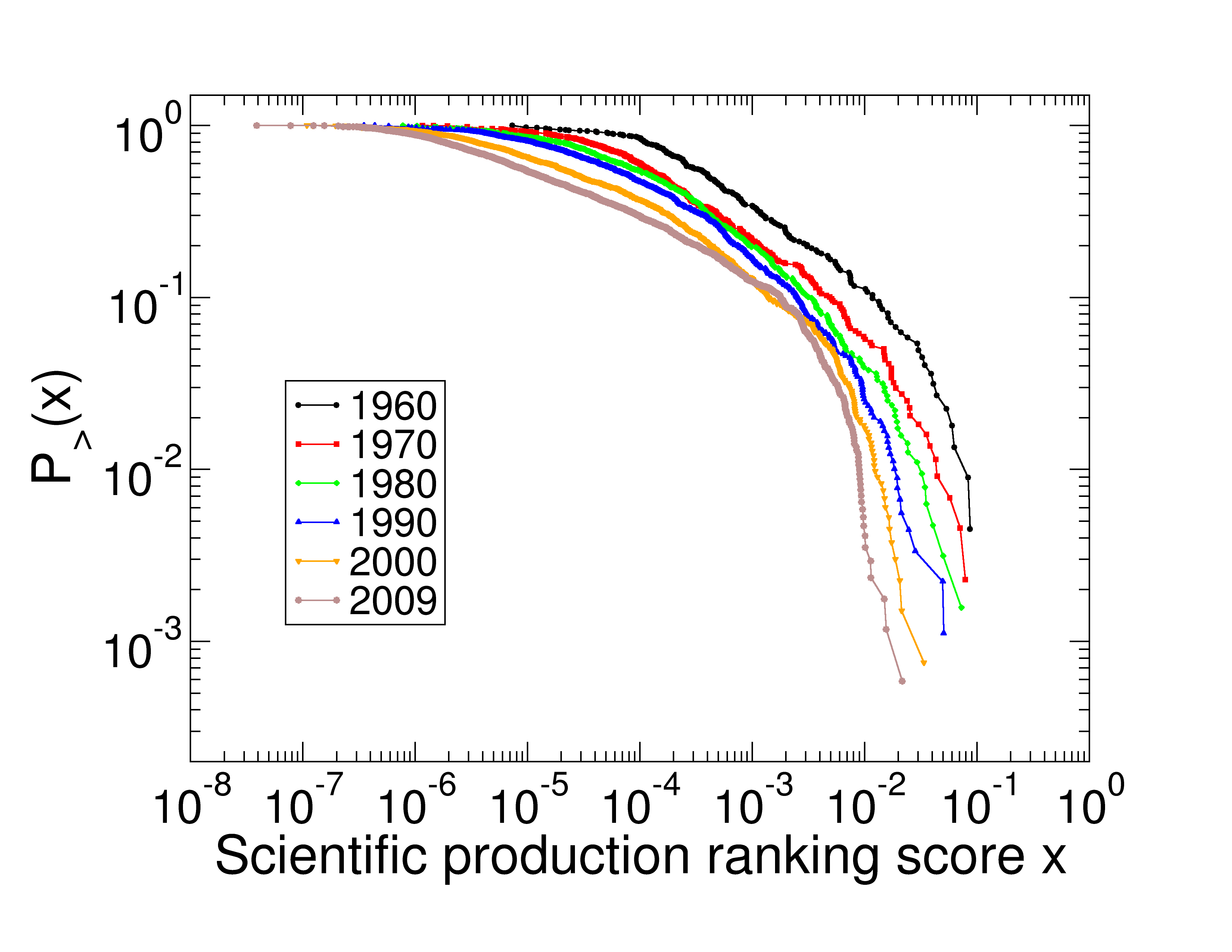}
\caption{The cumulative distribution function of scientific production ranking scores for cities in year 1960, 1970, 1980, 1990, 2000 and 2009.}
\label{fig:appendix:rankingdistribution}
\end{figure}

We show the cumulative distribution of scientific production ranking scores for cities in selected years in \autoref{fig:appendix:rankingdistribution}. We notice that ranking scores are also characterized with heavy tail distributions. In addition, we also observe that both the maximum and minimum ranking scores has decreased with time, and the tail of the distribution becomes steeper in recent decades, which indicates the differences of ranking scores between top ranked cities have gradually shrunk. \\

In \autoref{tb:rank:city:1985} and \autoref{tb:rank:city:1960}, we report top 50 cities ranked from scientific production ranking algorithm from 1985 to 2009 and from 1960 to 1980 respectively.
\begin{table}
\centering
\caption{Top 50 cities from scientific production ranking algorithm (1985-2009)}
\label{tb:rank:city:1985}
\scalebox{0.8}{
\begin{tabular}{l*{6}{l}}
\hline
rank & 1985 & 1990 & 1995 & 2000 & 2005 & 2009\\
\hline
1 & Piscataway & Piscataway & Boston & Boston & Boston & Boston\\
2 & Boston & Boston & Piscataway & Berkeley & Los Angeles & Berkeley\\
3 & Berkeley & Berkeley & Berkeley & Piscataway & Berkeley & Los Angeles\\
4 & Palo Alto & Palo Alto & Los Angeles & Los Angeles & Orsay & Tokyo\\
5 & New York City & Yorktown Heights & New York City & New York City & Tokyo & Orsay\\
6 & Los Angeles & Los Angeles & Urbana & Chicago & Princeton & Chicago\\
7 & Ithaca & New York City & Chicago & Urbana & Piscataway & Paris\\
8 & Los Alamos & Los Alamos & Lemont & Rochester & Palo Alto & Princeton\\
9 & Princeton & Princeton & Palo Alto & Batavia & New York City & Rome\\
10 & Yorktown Heights & Urbana & Batavia & West Lafayette & Philadelphia & Piscataway\\
11 & Lemont & Chicago & Philadelphia & Lemont & Urbana & London\\
12 & Urbana & Philadelphia & Madison & Orsay & Santa Barbara & Urbana\\
13 & Chicago & Ithaca & Rochester & East Lansing & Rome & Lemont\\
14 & Philadelphia & Lemont & West Lafayette & Ann Arbor & Columbus & Philadelphia\\
15 & Orsay & Orsay & Orsay & Tokyo & College Park & Oxford\\
16 & DC & Santa Barbara & Princeton & College Station & New Haven & Santa Barbara\\
17 & College Park & College Park & Los Alamos & Tsukuba & Lemont & New Haven\\
18 & Oak Ridge & Oak Ridge & Rome & Philadelphia & Madison & Rochester\\
19 & Santa Barbara & Livermore & Tsukuba & Palo Alto & Paris & Madison\\
20 & Rochester & Batavia & Santa Barbara & Madison & San Diego & Columbus\\
21 & Rehovot & Tokyo & Yorktown Heights & College Park & Chicago & College Park\\
22 & San Diego & Rochester & College Station & Pittsburgh & Tsukuba & Batavia\\
23 & Pittsburgh & San Diego & Pittsburgh & Rome & Oxford & Moscow\\
24 & New Haven & Columbus & Ithaca & Princeton & Oak Ridge & East Lansing\\
25 & Stony Brook & Madison & College Park & Los Alamos & Tallahassee & Palo Alto\\
26 & Seattle & Pittsburgh & New Haven & New Haven & Rochester & Pittsburgh\\
27 & Columbus & DC & Ann Arbor & Toyonaka & Beijing & San Diego\\
28 & Boulder & Rehovot & Pisa & Durham & Pittsburgh & Ann Arbor\\
29 & Paris & Stuttgart & Waltham & Columbus & Ames & Tsukuba\\
30 & Livermore & Paris & East Lansing & Stony Brook & West Lafayette & Seoul\\
31 & Madison & Minneapolis & Oak Ridge & Santa Barbara & Batavia & Pisa\\
32 & Austin & Boulder & Tokyo & Albuquerque & Pisa & West Lafayette\\
33 & Tokyo & New Haven & Stony Brook & Baltimore & Boulder & Padua\\
34 & J{\"u}lich & West Lafayette & San Diego & Toronto & Padua & Dubna\\
35 & Zurich & Stony Brook & Minneapolis & Pisa & London & Evanston\\
36 & Batavia & Bloomington & Baltimore & Tallahassee & Montreal & Ames\\
37 & Bloomington & Seattle & Padua & Waltham & Livermore & New York City\\
38 & Minneapolis & Ann Arbor & Toronto & Ithaca & Los Alamos & Toronto\\
39 & West Lafayette & Austin & Boulder & Moscow & Seoul & Oak Ridge\\
40 & Ann Arbor & Zurich & Albuquerque & Montreal & East Lansing & Baltimore\\
41 & East Lansing & Vancouver & Stuttgart & Padua & Moscow & Beijing\\
42 & Stuttgart & Holmdel & Livermore & San Diego & Nashville & Karlsruhe\\
43 & Evanston & Rome & DC & Ames & Ann Arbor & Taipei\\
44 & Grenoble & Ames & Paris & Evanston & College Station & College Station\\
45 & Syracuse & Waltham & Seattle & Meyrin & Vancouver & Meyrin\\
46 & Providence & Albuquerque & Rehovot & Gainesville & Irvine & Los Alamos\\
47 & Ames & Toyonaka & Durham & Honolulu & Taipei & Toyonaka\\
48 & Albany & Albany & Toyonaka & Paris & Dallas & Liverpool\\
49 & Waltham & J{\"u}lich & Columbus & Oak Ridge & Meyrin & Davis\\
50 & Nashville & Grenoble & Dallas & Bloomington & Cincinnati & Amsterdam\\
\hline
\end{tabular}
}
\end{table}
\begin{table}
\centering
\caption{Top 50 cities from scientific production ranking algorithm (1960-1980)}
\label{tb:rank:city:1960}
\scalebox{0.8}{
\begin{tabular}{l*{5}{l}}
\hline
rank & 1960 & 1965 & 1970 & 1975 & 1980\\
\hline
1 & Berkeley & Berkeley & Boston & Boston & Boston\\
2 & Boston & Boston & Berkeley & Piscataway & Piscataway\\
3 & New York City & Princeton & Piscataway & Berkeley & Berkeley\\
4 & Princeton & Piscataway & Palo Alto & Palo Alto & Palo Alto\\
5 & Chicago & New York City & Princeton & New York City & New York City\\
6 & Piscataway & Chicago & New York City & Princeton & Princeton\\
7 & Urbana & Los Angeles & Chicago & Ithaca & Los Angeles\\
8 & Los Angeles & Urbana & Los Angeles & Los Angeles & Chicago\\
9 & Ithaca & Palo Alto & Urbana & Chicago & Ithaca\\
10 & Pittsburgh & Pittsburgh & Ithaca & Lemont & Lemont\\
11 & Oak Ridge & Lemont & Pittsburgh & Urbana & Los Alamos\\
12 & Los Alamos & DC & Lemont & Batavia & Philadelphia\\
13 & DC & Ithaca & San Diego & Philadelphia & Urbana\\
14 & Rochester & Los Alamos & Oak Ridge & Oak Ridge & Oak Ridge\\
15 & Philadelphia & Albany & Philadelphia & Pittsburgh & College Park\\
16 & Albany & Oak Ridge & DC & College Park & Batavia\\
17 & Palo Alto & Philadelphia & Albany & DC & Orsay\\
18 & Lemont & Waltham & New Haven & San Diego & Stony Brook\\
19 & New Haven & New Haven & Waltham & Rochester & DC\\
20 & Madison & Madison & College Park & Los Alamos & Pittsburgh\\
21 & College Park & San Diego & Los Alamos & New Haven & Rochester\\
22 & Bloomington & College Park & Madison & Madison & Yorktown Heights\\
23 & Waltham & Rochester & Rochester & Waltham & New Haven\\
24 & Ann Arbor & Ann Arbor & Ann Arbor & Stony Brook & San Diego\\
25 & Minneapolis & Livermore & West Lafayette & Yorktown Heights & Rehovot\\
26 & West Lafayette & West Lafayette & Livermore & Albany & Madison\\
27 & Houston & Meyrin & Minneapolis & Orsay & Livermore\\
28 & Syracuse & Seattle & Rehovot & Seattle & Seattle\\
29 & Livermore & Minneapolis & Oxford & Providence & Waltham\\
30 & Columbus & Rehovot & London & Livermore & Albany\\
31 & Durham & Cleveland & Yorktown Heights & Rehovot & Evanston\\
32 & St Louis & Yorktown Heights & Meyrin & Minneapolis & West Lafayette\\
33 & Oxford & Oxford & Orsay & Evanston & Austin\\
34 & Cleveland & London & Ames & Durham & Providence\\
35 & Baltimore & Bloomington & Evanston & West Lafayette & Minneapolis\\
36 & Seattle & Evanston & Seattle & Ames & Ann Arbor\\
37 & Providence & Cambridge & Cleveland & London & Albuquerque\\
38 & Rehovot & St Louis & Stony Brook & Ann Arbor & Paris\\
39 & Ames & Syracuse & Cambridge & Cleveland & East Lansing\\
40 & Cambridge & Ames & Providence & East Lansing & Bloomington\\
41 & London & Detroit & Durham & Albuquerque & Cleveland\\
42 & Ottawa & Columbus & Santa Barbara & Austin & College Station\\
43 & Tokyo & Durham & Boulder & Oxford & Zurich\\
44 & Meyrin & Orsay & Riverside & Santa Barbara & Oxford\\
45 & Detroit & Houston & St Louis & St Louis & Ames\\
46 & South Bend & Boulder & Hamburg & Boulder & London\\
47 & Birmingham & Baltimore & Detroit & Columbus & Durham\\
48 & Jerusalem & Tokyo & Columbus & Zurich & Boulder\\
49 & San Diego & Paris & Syracuse & Cambridge & St Louis\\
50 & Sydney & Rome & Bloomington & Rome & Columbus\\
\hline
\end{tabular}
}
\end{table}

\clearpage
\section{Relation between research outputs and investment}
\begin{figure}[ht]
\centering
\includegraphics[width=0.8\textwidth]{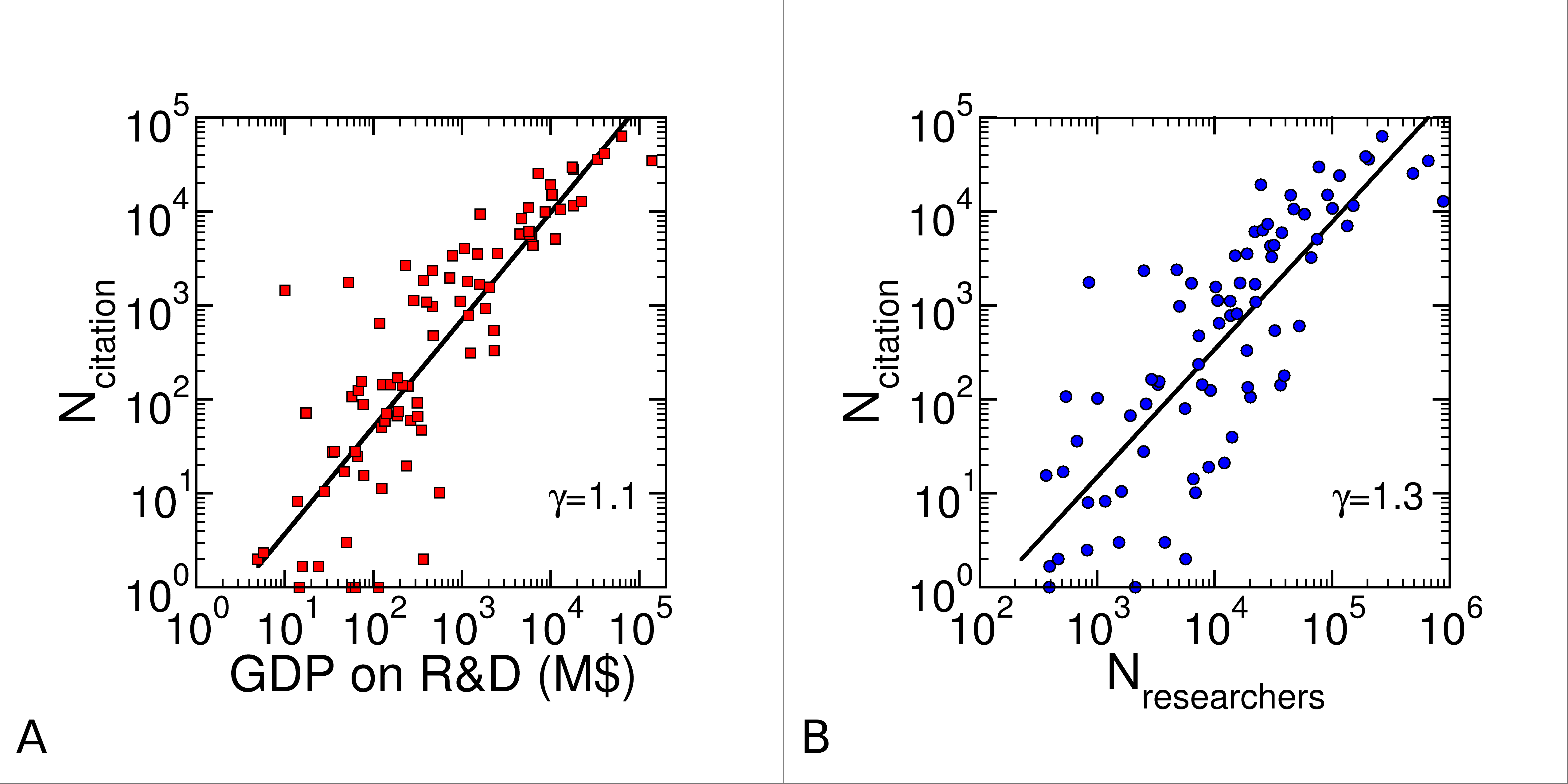}
\caption{Relation between research outputs and the investment. (A) The average citations received by each country as a function of the average GDP on research and development (R\& D) in million US dollars from $1996$ to $2009$. (B) The average citations received by each country as a function of the average research population in that country from $1996$ to $2009$. The solid black line shows the power-law fitting with the exponent $1.1$ and $1.3$ respectively.}
\label{fig:appendix:citationGDP}
\end{figure}

In this section, we report the relation between research outputs (i.e., citations) and investment on scientific research.  As discussed earlier, we parsed city information based on country information for each affiliation, therefore we can aggregate the number of citations for cities to their countries, and measure the relation between research outputs and investment on research in that country. In \autoref{fig:appendix:citationGDP}, we plot the correlation between the average number of citations received by each country in 1996-2009 and the average amount of gross domestic product (GDP) spent on research and development  (R\& D) (in current US dollars) in that country in that period. We also plot the correlation between the average number of citations received by one country in the same period and the average research population in that country within the same time window. The number of citations received approximately linearly scales with both quantities. Such findings are consistent with the results reported in \cite{Fortunato2012}, which studied the database of the Institute for Scientific Information (ISI). This similarity indicates, although APS dataset is limited, it is representative of the scientific production for major countries. The data of GDP, the fraction of GDP spent on R\& D, and the research population are from The World Bank data \cite{WorldBankData}. 
\end{appendices}

\end{document}